\documentclass[prx,aps,twocolumn,reprint,showpacs,floatfix,superscriptaddress,longbibliography]{revtex4-1}
\usepackage{graphicx}
\usepackage{amsmath}
\usepackage{ulem}
\usepackage{xcolor}
\usepackage{euscript}
\usepackage{mathrsfs}

\newcommand{\ket}[1]{|#1\rangle}
\newcommand{\bra}[1]{\langle#1|}

\newcommand{\ts}{\textsubscript}
\newcommand{\e}{\mathrm{e}}

\newcommand{\Tr}{\mathrm{Tr}}
\newcommand{\caO}{\mathcal O}
\newcommand{\caP}{\mathcal P}
\newcommand{\caL}{\mathcal L}

\newcommand{\caA}{\mathcal A}
\newcommand{\caY}{\mathcal Y}
\newcommand{\caK}{\mathcal K}

\definecolor{darkred}{rgb}{0.90,0.2,0.2}
\definecolor{darkgreen}{rgb}{0,0.60,.2}
\definecolor{darkblue}{rgb}{0.1,0.3,1}
\definecolor{grey}{cmyk}{0,0,0,0.25}
\definecolor{orange}{cmyk}{0,0.6,0.8,0}

\usepackage[plainpages=false,pdfpagelabels,colorlinks=true,linkcolor=red,urlcolor=blue,citecolor=blue,pdftitle={Quantum quenches and relaxation dynamics in the thermodynamic limit},pdfauthor={},pdfdisplaydoctitle=true,pdfduplex=DuplexFlipLongEdge]{hyperref}

\begin{document}

\title{Prethermalization and Thermalization in Isolated Quantum Systems}

\author{Krishnanand Mallayya}
\affiliation{Department of Physics, Pennsylvania State University, University Park, Pennsylvania 16802, USA}
\affiliation{Kavli Institute for Theoretical Physics, University of California, Santa Barbara, California 93106, USA}
\author{Marcos Rigol}
\affiliation{Department of Physics, Pennsylvania State University, University Park, Pennsylvania 16802, USA}
\affiliation{Kavli Institute for Theoretical Physics, University of California, Santa Barbara, California 93106, USA}
\author{Wojciech De Roeck}
\affiliation{Instituut voor Theoretische Fysica, KULeuven, 3001 Leuven, Belgium}
\affiliation{Kavli Institute for Theoretical Physics, University of California, Santa Barbara, California 93106, USA}

\date{\today}

\begin{abstract}
Prethermalization has been extensively studied in systems close to integrability. We propose a more general, yet conceptually simpler, setup for this phenomenon.  We consider a---possibly nonintegrable---reference dynamics, weakly perturbed so that the perturbation breaks at least one conservation law of the reference dynamics. We argue then that the evolution of the system proceeds via intermediate (generalized) equilibrium states of the reference dynamics. The motion on the manifold of equilibrium states is governed by an autonomous equation, flowing towards global equilibrium in a time of order $g^{-2}$, where $g$ is the perturbation strength.  We also describe the leading correction to the time-dependent reference equilibrium state, which is, in general, of order $g$. The theory is well confirmed in numerical calculations of model Hamiltonians, for which we use a numerical linked cluster expansion and full exact diagonalization.
\end{abstract}

\maketitle

\section{Introduction}

Prethermalization~\cite{berges2004prethermalization} has emerged over the past decade as an interesting and ubiquitous phenomenon in the dynamics of ultracold quantum gases in one-dimensional geometries~\cite{kinoshita2006quantum, gring2012relaxation, langen_erne_15, langen2016prethermalization, tang_kao_18}. In general, it refers to a separation of timescales: Some systems far from equilibrium quickly relax to long-lived (non)thermal states (not true thermal equilibrium states) on short timescales, before eventually relaxing to the expected true thermal equilibrium states on much longer timescales. 

There are some general instances of this phenomenon that are well understood, analytically and numerically. A first example is quenches in isolated noninteracting (integrable) systems, in which interactions (integrability-breaking perturbations) of strength $g$ are turned on~\cite{moeckel_kehrein_2008, *moeckel_kehrein_2009, eckstein_kollar_09, kollar_wolf_11, tavora_mitra_13, nessi_iucci_14, essler2014quench, bertini2015prethermalization, *bertini2016prethermalization, fagotti2015universal, reimann_dabelow_19}. By dephasing, observables quickly settle to quasisteady states that are well described by generalized Gibbs ensembles (GGEs)~\cite{rigol_dunjko_07, vidmar_rigol_16, essler_fagotti_16, cazalilla_chung_16, caux_review_16} of the noninteracting systems. The observables then relax to the thermal equilibrium values (thermalize) in a much longer timescale $\propto g^{-2}$, as predicted by kinetic Boltzmann-like equations. The latter can be derived applying time-dependent perturbation theory to the GGEs~\cite{stark_kollar_2013, dalessio_kafri_16} and, physically, describe the effects of collisions between (quasi)particles. It was recently shown numerically that weakly breaking integrability in strongly interacting integrable systems also results in thermalization rates $\propto g^{2}$~\cite{mallayya2018quantum}. A second example is isolated weakly interacting driven systems, for which GGEs and kinetic Boltzmann-like equations are also relevant~\cite{lazarides_das_14GGE, canovi2016stroboscopic}. A third example is periodically driven systems at high frequency. In general, they quickly reach a time-periodic state that can be identified as a Gibbs state corresponding to an effective Hamiltonian, before relaxing to the thermal (infinite-temperature~\cite{dalessio_rigol_14, lazarides_das_14}) state in a timescale that is exponentially long in the frequency of the drive~\cite{bukov2015universal, abanin2017effective, abanin2017rigorous, mori2016rigorous, KUWAHARA2016, machado_meyer_17}.

In this work, we argue that the first two examples above are instances of a more universal phenomenon, a phenomenon that occurs whenever an equilibrating dynamics (to a thermal- or GGE-like state) is weakly perturbed so that, at least, one of the conserved quantities in the original dynamics is no longer conserved in the weakly perturbed dynamics. A similar point of view was put forward in Refs.~\cite{lenarvcivc2018perturbative,lange2018time} for open quantum systems. We first present this conclusion in a loose manner in Sec.~\ref{sec:preamble}, and then, in Secs.~\ref{sec:slowdynamics} and~\ref{sec:generalobservables}, we develop a systematic treatment of weakly perturbed systems in which a conservation law is broken. The remainder of the paper is dedicated to demonstrating, in the context of numerical experiments, the validity and accuracy of this systematic treatment. Section~\ref{sec:intronum}, in which we introduce the models, quenches, observables, etc, is the preamble to the numerical experiments, which are reported in Sec.~\ref{sec:num}. A summary and discussion of our results is presented in Sec.~\ref{sec:summary}, while the Appendices report details of our analytical and numerical calculations.

\section{Preamble}\label{sec:preamble}

{\it Setup.}---We consider isolated extended quantum systems that are translationally invariant. For the sake of concreteness, one can think of chains with $L \gg 1$ sites. We have a pair of Hamiltonians $\hat H_0$ and $\hat H$, in which the latter includes the former plus a weak perturbation
\begin{equation}
\hat H=\hat H_0+g \hat V,
\end{equation}
where $g$ is small. We assume that the (reference) Hamiltonian $\hat H_0$ has a conservation law, say $\hat Q$, which is not shared by $\hat H$, namely, $[\hat H_0,\hat Q] =0$ but $ [\hat V,\hat Q] \neq 0$. We are interested in cases in which $\hat H_0$, $\hat Q$, $\hat H$, and $\hat V$ are extensive operators. In order to take the thermodynamic limit, it is helpful to deal with the intensive counterparts of those operators: $\hat h_0=\hat H_0/L$, $\hat q=\hat Q/L$, $\hat h=\hat H/L$, etc. In our numerical calculations, $\hat Q$ ($\hat q$) is the total number of particles (the filling), or, in the spin language, the total magnetization (the site magnetization).

{\it Assumption of fast equilibration.}---We consider cases in which the dynamics generated by $\hat H_0$, even when the system is far from equilibrium, results in fast equilibration. Namely, we assume that, for experimentally relevant translational-invariant initial states $\hat \rho_I$, observables converge within a time $\tau^*$ to the predictions of an ensemble (e.g., microcanonical) of statistical mechanics $\hat \rho_{e_0,q}$, with $(e_0,q)=(\langle \hat h_0 \rangle_{\hat\rho_{I}},\langle \hat q \rangle_{\hat\rho_{I}})$. The restriction ``experimentally relevant'' is put to avoid cases in which the initial state is a macroscopic superposition of states at different densities $(e_0,q)$ (see Ref.~\cite{rigol2008thermalization} for a discussion of this issue in the context of quantum quenches). The restriction to translation-invariant states is used to avoid having leading effects in the equilibration dynamics that are $L$ dependent, e.g., particle or energy transport, which would complicate the picture. 
 
{\it Initial prethermalization.}---Even though there is no standard perturbation theory around a genuinely interacting $\hat H_0$, it is reasonable to assume that the effect of the perturbation is small at times $\tau \ll 1/g$, in the sense that, for such times, one can meaningfully approximate the dynamics under $\hat H$ by the reference dynamics generated by $\hat H_0$.  Therefore, by the assumption of fast equilibration, one should observe a fast initial relaxation of observables toward the predictions of $\hat \rho_{e_0,q}$. The latter state can be very different from the thermal equilibrium ensemble $\hat \rho_e$ associated to $\hat H$. $\hat \rho_{e_0,q}$ differs from $\hat \rho_e$ by the fact that $q$ is an additional constraint [of course, it also differs because $e_0$ is the energy density of $\hat H_0$ and not that of $\hat H$, but that difference is only $\caO(g)$]. 
  
{\it Thermalization rate.}---To determine how fast the true equilibrium is approached, let us start from $\hat \rho_{e_0,q}$ and look at the change in $\langle \hat Q(\tau)\rangle_{{e_0,q}}$. We do this perturbatively:
\begin{equation}\label{eq:dg2}
\langle \hat Q(\tau)\rangle_{e_0,q}=  \langle \hat Q \rangle_{e_0,q} + ig \tau \langle [\hat V,\hat Q]\rangle_{e_0,q}+ \caO(g^2\tau \tau^*L)
\end{equation}
where we use that $[\hat H_0,\hat{Q}]= [\hat H_0,\hat \rho_{e_0,q}]=0$. The precise error estimate $\caO(\cdot)$ will only be argued for later. The important point to be highlighted from Eq.~\eqref{eq:dg2} is that the leading order correction to $ \langle \hat Q \rangle_{e_0,q}$ vanishes because $\langle [\hat V,\hat Q]\rangle_{e_0,q} =0$, by the cyclic property of the trace and the fact that $[\hat{Q},\hat \rho_{e_0,q}]=0$. This suggests that the thermalization rate is $\propto g^2$. Indeed, carrying out the expansion one order further, we recognize Fermi's golden rule. Finally, if we had replaced $\hat{Q}$ by a general observable $\hat{O}$, the first order term does not vanish. It results in a universal deviation of $\langle\hat{O}\rangle$ in the instantaneous state from that of $\hat \rho_{e_0,q}$.

\section{Slow dynamics of approximately conserved quantities}\label{sec:slowdynamics}

In this section, we present a derivation of an approximate autonomous equation governing the dynamics of the densities $(e_0,q)$. This is Eq.~\eqref{eq:driftequation}, or, at a more abstract level, Eq.~\eqref{eq:firsteomrecast}. Our derivation is not mathematically rigorous, it uses physical assumptions and goes substantially beyond the heuristics presented above. Its validity and accuracy are confirmed by numerical calculations in Sec.~\ref{sec:num}.

\subsection{Slow variables}\label{sec:slowvariables}
We identify the densities $(e_0,q)$ as slow variables. To set up a controlled derivation, we need a projection map from states $\hat\rho$ to $(e_0,q)$. It turns out, however, that it is more natural to start with a projection $\caP$ from $\hat\rho$ to a probability distribution $p$ on $(e_0,q)$.  Indeed, each such $p$ can be lifted to a $\hat \rho$ by
\begin{equation}\label{eq:densityfromdistribution}
\hat\rho_p =\int \text{d}e_0\text{d}q \, p(e_0,q) \hat\rho_{e_0,q},
\end{equation}
and the physically most natural case is, of course, when $p(e_0,q)$ is a Dirac delta distribution $\delta(e_0-e_0^*)\delta(q-q^*)$; cf.~the discussion on ``experimentally relevant states'' in Sec.~\ref{sec:preamble}. To construct the map $\caP$, a natural approach is to measure $\hat H_0$ and $\hat Q$: 
\begin{equation}
p_{\hat\rho}(e_0,q) \propto  \Tr \left[ \hat{\rho}\, P(\hat H_0 \approx {e}_0L) P( \hat Q \approx {q} L) \right],
\end{equation}
where $P(\hat H_0 \approx {e}_0 L)$ is a spectral projection of $\hat H_0$ on the interval $[{e}_0L-\delta_E,{e}_0L+\delta_E ]$ with a resolution $\delta_E$ that is much larger than the level spacing, but smaller than any relevant energy scale (like, e.g., the energy per site). For $P( \hat Q \approx {q} L)$, we can simply take the projection of $\hat Q$ on the eigenvalue closest to ${q} L$. Then we can set
\begin{equation}
\caP\hat{\rho}=\int \text{d} e_0 \text{d} q \, p_{\hat\rho}(e_0,q) \hat\rho_{e_0,q},
\end{equation}
where ``$\int \text{d} e_0 \text{d} q$'' represents a sum with the aforementioned resolution. 
We define $\bar\caP=1-\caP$, and note that both $\caP,\bar\caP$ are projectors, namely, $\caP^2=\caP$ and $\bar \caP^2=\bar \caP$. A warning is in order, despite the fact that $\caP \hat\rho$ and $\bar\caP \hat\rho$ encode all the information about the microscopic distributions of $E_0$ and $Q$ (whose widths are expected to be subextensive in $E_0$ and $Q$~\cite{rigol2008thermalization}), all that information is not needed and our analysis does not allow one to keep track of it in time. Our results only depend on, and keep track of, the distribution of densities $p(e_0,q)$.

The above definition of the projection $\caP$ is motivated by ensembles of statistical mechanics, which are expected to describe many-body quantum systems after equilibration. An equivalent definition can be motivated purely from quantum mechanics. If one takes $\hat \rho$ and evolves it under the unitary evolution dictated by $\hat H_0$, observables equilibrate to the predictions of the so-called diagonal ensemble $\hat\rho_{\mathrm{DE}}$~\cite{dalessio_kafri_16}: 
\begin{equation}\label{eq:caPrho}
\hat \rho_{\mathrm{DE}} = \sum_{E^0} (\langle E^0|\hat \rho | E^0 \rangle) |E^0\rangle \langle E^0|,
\end{equation}
where $|E^0\rangle$ are the eigenkets $\hat H_0$ and $\hat Q$, and we have assumed that either there are no degeneracies in the many-body energy spectrum or that, if present, they are unimportant. This is generically the case in interacting many-body quantum systems~\cite{dalessio_kafri_16}.

The map $\hat \rho\to \hat \rho_{\mathrm{DE}}$ is a projection as well; let us call it $\caP_{\mathrm{DE}}$. It provides an alternative path to identifying the slow variables. This is the case because, in recent years, we have come to realize that the ensembles defined by $\caP\hat\rho$ and $\caP_{\mathrm{DE}} \hat \rho$ are equivalent when it comes to their predictions for observables (few-body or local operators) in large systems. This is a consequence of eigenstate thermalization for nonintegrable (quantum chaotic) systems~\cite{deutsch1991quantum, srednicki1994chaos, rigol2008thermalization}, and generalized eigenstate thermalization for integrable systems~\cite{cassidy_clark_11, caux_essler_13, vidmar_rigol_16, caux_review_16,mori18}. Hence, it does not really matter whether one uses the projection $\caP$ or $\caP_{\mathrm{DE}}$, so we do not actively distinguish between them in most of our analytical derivations. In our numerical calculations we use $\caP_{\mathrm{DE}}$, as $\hat \rho_{\mathrm{DE}}$ can be calculated exactly in the same way for nonintegrable and integrable systems. $\caP$, on the other hand, demands that one identifies beforehand the relevant conserved quantities (not necessary for $\caP_{\mathrm{DE}}$). Keeping all this is mind, in what follows we do not actively distinguish between nonintegrable systems, for which the number of conserved quantities in the thermodynamic limit is $\caO(1)$, from integrable systems, for which the number of conserved quantities in the thermodynamic limit is infinite.

\subsection{Mori-Zwanzig approach}\label{sec:nakajima}
 
Let us introduce the Liouville superoperator $\caL=-i[\hat H,\cdot]$. Then, following Mori-Zwanzig \cite{zwanzig1960ensemble,mori1965transport}, see Appendix~\ref{app:nakajima-zwanzig}, the theory of linear ordinary differential equations gives us the following rewriting of the $\caP$-projected Liouville equation $\partial_\tau \hat\rho(\tau) = \caL \hat\rho(\tau)$: 
\begin{eqnarray}\label{eq:eomforp}
\partial_\tau\caP\hat\rho(\tau)&=& \caP\caL\caP\hat\rho(\tau)\\ &&+ \int_0^\tau \text{d} s \caP\caL e^{s \bar\caP \caL } \caL\caP\hat\rho(\tau-s) + \caP\caL e^{\tau \bar\caP \caL }\bar\caP\hat\rho_I,\nonumber
\end{eqnarray}
where $\hat\rho_I\equiv\hat\rho(\tau=0)$, and we use that the projector $\caP$ is time independent. To bring some structure to this equation, we now split 
\begin{equation}
\caL=\caL_0+\caL_1,\ \ \text{with} \ \ \caL_0=-i[\hat H_0,\cdot], \caL_1= -ig[\hat V,\cdot],
\end{equation}
and note the properties
\begin{equation}
\caL_0 \caP=\caP \caL_0=0,\quad  \caP\caL_1\caP=0.
\end{equation}
They follow from elementary considerations. This allows one to recast the above equation of motion as
\begin{equation}\label{eq:firsteom}
\partial_\tau\caP\hat\rho(\tau)=\int_0^\tau \text{d} s \caA_{s}\caP\hat\rho(\tau-s) + \caY_{\tau} \bar\caP\hat\rho_I,
\end{equation}
where we introduce
\begin{equation}\label{eq:caa}
\caA_{s}= \caP\caL_1 e^{s(\caL_0+ \bar\caP \caL_1) } \caL_1\caP\ \ \text{and}\ \
\caY_{\tau}= \caP\caL_1  e^{\tau(\caL_0+ \bar\caP \caL_1) }\bar\caP.
\end{equation}
The object $\caA_s$ represents a memory kernel. For completeness, we also report the expression for $\bar\caP\hat\rho(\tau)$ (see Appendix~\ref{app:nakajima-zwanzig}):
\begin{equation}\label{eq:eomforbarp}
\bar\caP\hat\rho(\tau)=e^{\tau\bar\caP \caL }\bar\caP\hat\rho_I + \int_0^\tau \text{d} s \bar\caP e^{s \bar\caP \caL } \caL\caP\hat\rho(\tau-s).
\end{equation}
It should be stressed that all the previous equations are exact, and, hence, they are not particularly useful. In the next section, we make some motivated approximations.

\subsection{Equilibration}

Our only assumption is that the dynamics generated by $\hat H_0$ results in fast equilibration of observables to the equilibrium state that is characterized by the expectation value of $\hat H_0$, $E_0=\text{Tr}(\hat{\rho}_I\hat H_0)$, and of $\hat Q$, $Q = \text{Tr}(\hat{\rho}_I\hat Q)$. Once again, we stress that we do not distinguish between nonintegrable reference Hamiltonians $\hat H_0$, for which observables equilibrate to the thermal predictions (thermalize~\cite{dalessio_kafri_16}), from integrable reference Hamiltonians, for which observables equilibrate to the GGE predictions (exhibit generalized thermalization~\cite{vidmar_rigol_16}). For our purposes, it does not matter whether equilibration is towards a traditional ensemble of statistical mechanics or towards a GGE: in both cases it means that, for generic (few-body) observables $\hat O$ and initial states $\hat\rho_I$,
\begin{equation}
\frac{|\Tr [\hat\rho_I \hat O_0(\tau)]- \Tr[\hat\rho_{e_0,q}\hat O]| }{|\Tr[\hat\rho_{e_0,q}\hat O]|} =: f(\tau)  \to 0,
\end{equation}
where $\hat O_0(\tau)=e^{i \tau \hat H_0}\hat O e^{-i \tau \hat H_0}$. We define an equilibration timescale $\tau^*(\epsilon)$, which is understood as the time $\tau$ at which $f(\tau)\approx \epsilon$ for some small dimensionless $\epsilon$. By fast equilibration, we mean that $\tau^*(\epsilon)\sim\caO(1)$ in the relevant time units of the problem. In Sec.~\ref{sec:num}, numerical simulations of a translationally invariant one-dimensional Hamiltonian indeed show that, in times that are $\caO(1)$ in the relevant timescale, observables reach values that are very close to the predictions of $\hat \rho_{e_0,q}$.

We do not expect $\epsilon$ to be arbitrarily small because, even if the Hamiltonian and the initial state are translational invariant, the unavoidable coupling to hydrodynamic modes in interacting systems renders the approach of local observables to equilibrium polynomial, $f(\tau) \propto \tau^{-d/2}$, as argued in Ref.~\cite{lux2014hydrodynamic} on the basis of fluctuating hydrodynamics. In such a case, one expects
\begin{equation}
\tau^*(\epsilon) \approx \epsilon^{-2/d}.
\end{equation}
Our numerical results suggest that, for the models and observables studied, hydrodynamics tails may set in when $\epsilon$ is very small, at times that are beyond the reach of our numerical calculations. 

The way $\tau^*$ enters our analysis is that we approximate, for any $\hat\rho$,
\begin{equation}\label{eq:thermalizationassumption}
e^{\tau\caL_0}\hat\rho\approx\caP e^{\tau\caL_0}\hat\rho, \qquad \text{whenever $\tau>\tau^*$},
\end{equation}
accepting an error $\caO(\epsilon)$. This approximation amounts to assuming that the system has equilibrated after a time $\tau^*$ with respect to the dynamics generated by $\hat H_0$. Hence, at time $\tau^*$, we replace the density matrix of the system by $\hat\rho_{e_0,q}$. Of course, the usual caveats typical of irreversibility apply: this replacement can only be correct when dealing with local or few-body observables $\hat{O}$, or sums thereof, not for (special) many-body operators such as the spectral projectors of $\hat{H}_0$~\cite{dalessio_kafri_16}. 

\subsection{Born approximation}\label{sec:born}

We can now state our main assumption as a weak coupling condition, namely,
\begin{equation}\label{eq:basicwccondition}
g\tau^* \ll 1.
\end{equation}
To see this assumption at work, let us expand the exponential $e^{s(\caL_0+ \bar\caP \caL_1)}$ in the definition of $\caA_s$ in Eq.~\eqref{eq:caa}
\begin{eqnarray}
\caA_s&=& \caP\caL_1 e^{s\caL_0} \caL_1\caP  \\&&+ \int_0^s \text{d}s_1 \caP\caL_1 e^{(s-s_1)\caL_0} \bar\caP \caL_1 e^{s_1\caL_0} \caL_1\caP + \cdots . \nonumber
\end{eqnarray}
Note that, for example, the first term on the rhs could also be written as $\caP\caL_1\bar\caP e^{s\caL_0} \bar\caP\caL_1\caP$ or even $\caP\caL_1\bar\caP e^{s\bar\caP\caL_0\bar\caP}  \bar\caP\caL_1\caP$, which makes apparent that the intermediate evolution acts on the fast degrees of freedom. In order for this expansion in powers of $g$ to be meaningful, one needs to make sure that the series above can be resummed such that it is linear in $L$. For example, since every $\caL_1$ carries a factor $L$, it appears that the first and second terms in the equation above are of order $L^2$ and $L^3$, respectively. However, one can write these terms as sums of [and integrals over $(e_0,q)$] truncated correlation functions $\langle \hat V_0(s)\hat V \rangle^{c}_{e_0,q}$ and $\langle \hat V_0(s)\hat V_0(s_1) \hat V \rangle^{c}_{e_0,q}$; see, e.g., the remark following Eq.~\eqref{eq:expressionloss}. These truncated correlation functions are of order $L$ due to clustering properties of the $(e_0,q)$ equilibrium ensembles. In higher orders, such considerations become more complicated and we refer the interested reader to mathematical work establishing a meaningful expansion~\cite{davies1974markovian, de2013diffusion}. Assuming that one recovers the linearity in $L$ to all orders, for any $\tau\geq \tau^*$,
\begin{equation}\label{eq:incaA}
\int_0^\tau \text{d}s \caA_s = \int_0^\tau \text{d}s\caP\caL_1 e^{s\caL_0} \caL_1\caP +  \caO(g^3 \tau^{*2}L),
\end{equation}
where the first term is $\caO(g^2L\tau^*)$. One then sees that the expansion is meaningful if 
\begin{equation}
g^3 \tau^{*2}L \ll g^2\tau^* L,
\end{equation}
which is, of course, equivalent to Eq.~\eqref{eq:basicwccondition}.

\subsection{Markov approximation}\label{sec:markov}

The discussion above has shown that the lowest order approximation to $\int_0^\infty d s \caA_s$, namely, 
\begin{equation}\label{eq:caK}
\caK = \int_0^\infty \text{d}s\caP\caL_1 e^{s\caL_0} \caL_1\caP,
\end{equation}
is $\caO(g^2\tau^* L)$, with the factor $L$ originating from the spatial sum in $\caL_1$. The physical meaning of this factor is that the quantities that change smoothly in time are the densities $(e_0,q)$ rather than $(E_0,Q)$ themselves, or, alternatively, local observables. We then see that the superoperator $\caK$ is $\caO(g^2\tau^*)$, and this is the rate at which $\hat\rho$ changes. Recalling that the time integral defining $\caK$ reaches its $\tau=\infty$ value at $\tau\approx\tau^*$ leads to the conclusion that, in Eq.~\eqref{eq:firsteom}, we can approximate $\caP \hat\rho(\tau-s)$ by $\caP\hat\rho(\tau)$, making an error $\caO(g^2\tau^{*2})$ (which is small by our weak-coupling assumption $g\tau^* \ll 1$). A further simplification occurs by observing that $\caY_\tau$ is proportional to $g$, and vanishes fast as $\tau\geq \tau^*$, so one might set $\caY_\tau=0$ for times $\tau \geq \tau^*$. If we make these two approximations, then the equation of motion [Eq.~\eqref{eq:firsteom}] reads
\begin{equation}\label{eq:firsteomrecast}
\partial_\tau\caP\hat\rho(\tau)= \caK \caP\hat\rho(\tau),\qquad \tau \gg \tau^*.
\end{equation}
A conservative look at the validity of Eq.~\eqref{eq:firsteomrecast}, in particular to the approximation $\caY_\tau =0$, shows that one should trust Eq.~\eqref{eq:firsteomrecast} only for times $\tau\gg \tau_{\mathrm{tr}}$  (where the subindex $\mathrm{tr}$ stands for transient) and $\tau_{\mathrm{tr}}$ is determined by $g^2\tau^* \gg  g f(\tau_{\mathrm{tr}})$, as follows by comparing the magnitude of the two terms in Eq.~\eqref{eq:firsteom}.

\subsection{Autonomous equation}\label{sec:autonomousequation}

The meaning of Eq.~\eqref{eq:firsteomrecast}, in which $\caK$ is a superoperator acting on density matrices, is simplified by the presence of $\caP$. As mentioned before, $\caP$ projects onto distributions of densities $p(e_0,q)$, and so Eq.~\eqref{eq:firsteomrecast} is actually an evolution equation on such $p$. It is easier, and more natural, to guess the form of this evolution equation than to derive it from the formula above, so we will do the former here and relegate the latter to Appendix~\ref{app:fgrfromcak}.

Given $\hat \rho_{e_0,q}$, we need to find the rate of change of $(e_0,q)$. The natural path is to use Fermi's golden rule, within which $e_0$ does not change in time. Having a single eigenket $|E^0\rangle$ in mind, the rate of change of $q$, called the ``drift'' $d$, is given by  
\begin{equation}\label{eq:drate}
d(E^0)=2\pi g^2 \int \text{d} \overline E^0 \delta (E^0-\overline E^0) \frac{Q_{\overline{E}^0}-Q^{}_{E^0}}{L} |\langle E^0|\hat V|\overline E^0\rangle|^2,
\end{equation}
where $Q^{}_{E^0}=\langle E^0|\hat Q|E^0\rangle$, the integral is understood as a sum, and the delta function $\delta (E^0-\overline E^0)$ selects an interval of energies with a width smaller than any relevant energy scale but much larger than the level spacing. In principle, one should average $d(E^0)$ over $E^0$, with the distribution provided by statistical mechanics $\hat\rho_{e_0,q}$, or by quantum mechanics, i.e., the diagonal ensemble $\hat\rho_{\mathrm{DE}}$. However, once again because of eigenstate thermalization (or its generalized version for integrable systems), we expect the average to be unnecessary for large $L$.  Hence, the drift  $d=d(e_0,q)$ does not depend on $|E^0\rangle$ but only on $(e_0,q)$. Even though $\hat{V}=\caO(L)$, the drift $d(e_0,q)$ is $\caO(1)$ because of decay of spatial correlations, see Appendix \ref{app:fgrfromcak}. We then obtain the autonomous equation
\begin{equation}\label{eq:driftequation}
\partial_\tau q(\tau) = d[e_0(\tau),q(\tau)],\qquad \partial_\tau e_0(\tau) = 0.
\end{equation}
The corresponding evolution equation for the distributions $p_\tau(e_0,q)$ is then
\begin{equation}\label{eq:evolutionondensities}
\partial_\tau p_\tau(e_0,q) = -d[e_0,q]\partial_{q}p_\tau(e_0,q),
\end{equation}
which lifts naturally to an evolution on $\caP\hat \rho$. The stationary solution of Eq.~\eqref{eq:driftequation} as $\tau\to\infty$ is denoted by $(e_0,q^*)$, where $q^*=q^*(e_0)$ is determined by 
\begin{equation}
d[e_0,q^*(e_0)]=0.
\end{equation}
Within Fermi's golden rule [Eq.~\eqref{eq:drate}], this is recognized as a detailed balance condition, indicating that $q^*(e_0)$ is the equilibrium value of $q$ given $e_0$. In other words, $q^*$ is determined by maximizing the entropy at fixed $e_0$ and, hence, the resulting ensemble $\hat \rho_{e_0,q^*}$ is equivalent to one in which no constraint on $\hat{q}$ is imposed:  
\begin{equation}
\hat \rho_{e_0,q^*}\approx \hat \rho_{e_0}.
\end{equation}
This also shows that the asymptotic state $\hat \rho_{e_0,q^*}$ is close to the global equilibrium state $\hat \rho_{e}$ for which $\langle\hat h\rangle=e=e_0+\caO(g)$. In Sec.~\ref{sec:generalobservables}, we quantify the $\caO(g)$ difference between $\hat \rho_{e_0,q^*}$ and $\hat \rho_{e}$.

\subsection{Corrections}

In principle, our scheme allows one to compute higher-order corrections in $g$ to the autonomous Eqs.~\eqref{eq:firsteomrecast} and~\eqref{eq:driftequation}. In particular, one can expand $\caA_s$ in a power series in terms of order $(g\tau^*)^nL$, $n\geq 1$, with the terms for $n=1$ and 2 given in Sec.~\ref{sec:born}. Integrating over $s$, one then gets a series for $\caK$, and hence for the drift coefficient $d[e_0,q]$. It is, however, far from obvious that such corrections are useful, as (i) we have made approximations at several points, not only in truncating $\caA_s$, but also, e.g., replacing $\caP\hat\rho(\tau-s)$ for $s\leq \tau^*$ by $\caP\hat\rho(\tau)$, and (ii) a not-so-fast decay of $f(\tau)$ would imply that the $\caY_\tau$ term in Eq.~\eqref{eq:firsteom} remains potentially important at long times, obscuring any precise corrections to the Markovian part. 

Rigorous work in the context of open quantum systems \cite{davies1974markovian, lukkarinen2007kinetic,de2013diffusion} implies that one can meaningfully compute corrections, but only when $\int_0^\infty \text{d} \tau|f(\tau)| <\infty$.

\section{Dynamics of local observables}\label{sec:generalobservables}
  
So far, we have focused on the evolution of the quasiconserved quantity $\hat{Q}$, which we found to be described by the autonomous Eq.~\eqref{eq:driftequation}. We now turn our attention to more general observables. Our main finding is that their time-dependent expectation values are described by a ``deformed equilibrium ensemble,'' with an $\caO(g)$ difference from $\hat\rho_{e_0,q(\tau)}$. This deviation has a universal form.   
  
\subsection{Correction to $\caP\hat\rho(\tau)$}\label{eq:correctionstocaprho}

At first sight, the evolution of generic observables is slaved by the evolution of the slow variables $(e_0,q)$. Indeed, the general picture is that the evolution of the density matrix $\hat\rho(\tau)$ takes place in the space of equilibrium density matrices $\hat\rho_{e_0,q(\tau)}$, leading to the prediction
\begin{equation}\label{eq:oslaved}
\Tr[\hat\rho(\tau) \hat O] \approx \Tr [\hat O\caP\hat\rho(\tau)] = \langle \hat O \rangle_{e_0,q(\tau)}.
\end{equation}
In the last equality, we assumed that $p(e_0,q)$ is concentrated at a single value. Yet, strictly speaking, the previous section dealt with $\caP\hat\rho(\tau)$, rather than with $\hat\rho(\tau)$, so Eq.~\eqref{eq:oslaved} is in need of a justification. To provide it, we return to the formalism in Sec.~\ref{sec:nakajima}, and write [see Eq.~\eqref{eq:eomforbarp}]
\begin{eqnarray}\label{eq:justO}
\hat \rho(\tau) &=& \caP\hat\rho(\tau) +  \bar\caP\hat\rho(\tau) \\
&=& \caP\hat\rho(\tau) +  \e^{\tau \bar\caP \caL }\bar\caP\hat\rho_I + \int_0^\tau \text{d} s \bar\caP e^{s \bar\caP \caL} \caL\caP\hat \rho(\tau-s).\nonumber 
\end{eqnarray}

The first term in Eq.~\eqref{eq:justO} is dominant for large $\tau\geq \tau_{\mathrm{tr}}$. It amounts to the approximation resulting in Eq.~\eqref{eq:oslaved}. The second term in the last line of Eq.~\eqref{eq:justO} is transient, decaying as $gf(\tau)$, as discussed also in Sec.~\ref{sec:markov}. The third term is the correction we are interested in. Let us write it explicitly, when paired with an observable $\hat O$, namely, $\int_0^\tau \text{d} s\, \Tr[\hat O\bar\caP e^{s \bar\caP \caL} \caL\caP\hat \rho(\tau-s)]$. By expanding $\e^{s\bar\caP \caL}$ in powers of $g$, we get the leading contribution
\begin{eqnarray}\label{eq:just1}
&\int_0^\tau \text{d} s&\, \Tr [\hat O \bar\caP e^{s \caL_0 }\caL\caP\hat\rho(\tau-s)]\nonumber\\&& = {i g\int_0^{\tau} \text{d} s \, \Tr\{[\hat V_0(-s),\hat O] \caP\hat\rho(\tau-s)\}.}
\end{eqnarray}
One can further simplify this expression by approximating $\caP\hat\rho(\tau-s)$ by $\caP\hat\rho(\tau)$, justified by the reasoning in Sec.~\ref{sec:markov}, and by again assuming that the distribution $p(e_0,q)$, corresponding to $\caP\hat\rho(\tau)$, is concentrated at a single density $[e_0,q(\tau)]$. Then the last line in Eq.~\eqref{eq:just1} reads
\begin{equation}\label{eq:correctiontoplateau}
{ig \int_0^{\infty} \text{d} s \langle [\hat V_0(-s),\hat O] \rangle_{e_0,q(\tau)}}.
\end{equation}
By the assumption of fast equilibration at $s\approx \tau^*$, this expression is $\caO(g\tau^*)$; i.e., it is a small correction to $\langle \hat O \rangle_{e_0,q(\tau)}$. Even though it is subleading, this correction has a universal form, as we explain in the next section. Finally, note that Eq.~\eqref{eq:correctiontoplateau}, obtained following a naive first-order perturbation theory, was already written in Sec.~\ref{sec:preamble} for $\hat{Q}$ [see Eq.~\eqref{eq:dg2}]. In that case, the $s$-independent integrand vanishes identically.

\subsection{Susceptibility}\label{sec:susceptibility}

To understand the correction in Eq.~\eqref{eq:correctiontoplateau}, consider the equilibrium ensemble $\hat\rho_{e_0,q^*}$ with $(e_0,q^*)$ the $\tau\to\infty$ solution of Eq.~\eqref{eq:driftequation}, or, alternatively, with $q^*$ determined by maximizing the entropy at fixed $e_0$. The ensemble $\hat\rho_{e_0,q^*}$ is a small perturbation of the real equilibrium ensemble $\hat\rho_e$ corresponding to $\hat H$, with $\langle \hat h \rangle_{\hat\rho_e}=\langle \hat h \rangle_{\hat\rho_{e_0,q^*}} + \caO(g)$. We can relate these two ensembles by linear response theory. Indeed, we can imagine starting with $\hat\rho_{e_0,q^*}$ at $\tau=0$, and switching on the perturbation $g\hat V$. The system will then evolve precisely to $\hat\rho_e$, and the change in the expectation value of observables $\hat O$ should be described by linear response theory. In particular, the stationary change is described by the zero-frequency response coefficient (also known as the susceptibility), which is exactly Eq.~\eqref{eq:correctiontoplateau}. 

This discussion should also clarify how thermalization with respect to $\hat H$ is reconciled with our treatment, which is based on equilibration with respect to $\hat H_0$. The global equilibrium state $\hat{\rho}_e$ is obtained as a universal correction to the state $\hat{\rho}_{e_0,q^*}$.

\subsection{Deformed equilibrium ensembles from Fermi's golden rule}\label{sec:deformedgge}

In Sec.~\ref{eq:correctionstocaprho}, we use the last term in Eq.~\eqref{eq:justO} to derive the $\caO(g)$ correction to observables in $\hat{\rho}(\tau)$ from their expectation values in $\hat \rho_{e_0,q(\tau)}$. Here, we point out that this is also what the autonomous Eq.~\eqref{eq:driftequation} dictates.

Let us compute the time derivative of $q(\tau)$:
\begin{equation}\label{eq:derivativedifferently}
\partial_\tau q(\tau)= \Tr [\hat{q} \caL\hat\rho(\tau)]= \Tr [\hat{q} \caL_1\hat \rho(\tau)],
\end{equation}
where, in the last equality, we use that $[\hat{H}_0,\hat Q]=0$. Previously, we evaluated this time derivative in a different way: we approximated $\hat\rho(\tau)$ by $\hat\rho_{e_0,q(\tau)}$ [assuming, again, that $p(e_0,q)$ is a Dirac delta function] and we arrived at
\begin{equation}\label{eq:derivativeoriginally}
\partial_\tau q(\tau) = \Tr [\hat q \caK \hat\rho_{e_0,q(\tau)}].
\end{equation}
This is a more abstract rendering of the autonomous Eq.~\eqref{eq:driftequation}. At this point, one can ask which form of $\hat\rho(\tau)$ would reconcile Eqs.~\eqref{eq:derivativedifferently} and~\eqref{eq:derivativeoriginally}. By the cyclic property of the trace, and using that $[\hat{q},\hat\rho_{e_0,q(\tau)}]=0$, we note that Eq.~\eqref{eq:derivativedifferently} does not depend on the leading contribution $\hat\rho_{e_0,q(\tau)}$ to $\hat{\rho}(\tau)$. It depends only on the correction term. We then see that Eq.~\eqref{eq:derivativedifferently} reduces to Eq.~\eqref{eq:derivativeoriginally} if we choose the correction term to be
\begin{equation}
\int_0^{\infty} \text{d}s \bar\caP\e^{s\caL_0} \caL_1 \hat\rho_{e_0,q(\tau)},
\end{equation}
which is the correction we used in Sec.~\ref{eq:correctionstocaprho} to derive Eq.~\eqref{eq:correctiontoplateau}. A spectacular corollary of this reasoning is the observation that, if $\hat\rho(\tau)$ were exactly commuting with $\hat{q}$, then the rate of change vanishes, i.e., $\partial_\tau q(\tau)=0$.

\section{Models, Quenches, and Observables}\label{sec:intronum}

In the rest of the paper, we test the previous ideas and analytical results in numerical experiments. We focus on the dynamics of nonintegrable models of strongly interacting hard-core bosons in one-dimensional lattices (which can be mapped onto spin-1/2 Hamiltonians). We study the effects of breaking particle-number conservation [the U(1) symmetry in the spin models]. 

\subsection{Models}

Quantum dynamics are studied under time-independent Hamiltonians $\hat{H}_\alpha$ of the form
\begin{equation}\label{H_f}
\hat{H}_\alpha=\hat{H}_0+g_\alpha\hat{V}_\alpha,
\end{equation}
where the reference Hamiltonian $\hat{H}_0$, which we take to be nonintegrable, commutes with the total particle-number operator $\hat N$, $[\hat{H_0}, \hat{N}]=0$. The perturbations, $g_\alpha\hat{V}_\alpha$, do not commute with $\hat N$, $[\hat{V}_\alpha, \hat{N}]\ne0$.

We take $\hat{H}_0$ to be the $t$-$V$-$t'$-$V'$ Hamiltonian for hard-core bosons in 1D lattices~\cite{rigol2009breakdown, santos2010onset}, with nearest (next-nearest) neighbor hopping $t$ ($t'$) and interaction $V$ ($V'$):
{\setlength\arraycolsep{0.1pt}
\begin{eqnarray}
&&\hat{H_0}=\sum_i \left[ -t\left( \hat{b}^\dagger_i \hat{b}^{}_{i+1} + \textrm{H.c.} \right) -t'\left( \hat{b}^\dagger_i \hat{b}^{}_{i+2} + \textrm{H.c.} \right) 
\right.\label{model_H}\\
&&\left.+V\left(\hat{n}^{}_i-\dfrac{1}{2}\right)\hspace*{-0.1cm}\left(\hat{n}^{}_{i+1}-\dfrac{1}{2}\right)+V'\left(\hat{n}^{}_i-\dfrac{1}{2}\right)\hspace*{-0.1cm}\left(\hat{n}^{}_{i+2}-\dfrac{1}{2}\right)\hspace*{-0.05cm}\right],\nonumber
\end{eqnarray}
}where $\hat{b}^\dag_i$ $(\hat{b}^{}_i)$ is the hard-core boson creation (annihilation) operator and $\hat{n}^{}_i=\hat{b}^\dag_i\hat{b}^{}_i$ is the number operator at site $i$. When $t'=V'=0$, $\hat{H_0}$ is integrable (in the spin language, it is the Hamiltonian of the spin-1/2 XXZ chain)~\cite{cazalilla_citro_review_11}. Here, we focus on cases in which $t'=V'\neq0$, so that $\hat{H}_0$ is nonintegrable~\cite{rigol2009breakdown, santos2010onset}.

We consider two perturbations $g_\alpha\hat{V}_\alpha$, with $\alpha=1,2$. The first one is
\begin{equation}
g_1\hat{V}_1 = g_1\sum_i\left[\hat{b}^{}_{i}+\frac{1}{2}\left(\hat{b}^{}_{i}\hat{b}^{}_{i+1} - \hat{b}^{\dagger}_{i}\hat{b}^{}_{i+1}\right)+ \text{H.c.}\right].\label{V1}
\end{equation}
It will be important later that the presence of nearest neighbor hopping terms in $\hat{V}_1$ make $\bra{E_i^0}\hat{V}_1\ket{E_i^0}\ne0$ for typical eigenkets $\ket{E^0_i}$ of $\hat{H}_0$.

The second perturbation we consider is
\begin{equation}
g_2\hat{V}_2=g_2\sum_i\left(\hat{b}^{}_{i}+\frac{1}{2}\hat{b}^{}_{i}\hat{b}^{}_{i+1} + \text{H.c.}\right).\label{V2}
\end{equation}
This perturbation only contains terms that change the particle number. Hence, $\bra{E_i^0}\hat{V}_2\ket{E_i^0}=0$ for all eigenkets $\ket{E^0_i}$ of $\hat{H}_0$ and $\hat{N}$.

\subsection{Initial states and description after equilibration}

We study the quantum dynamics of initial states $\hat{\rho}_I$ that are far from equilibrium with respect to both $\hat{H}_0$ and $\hat{H}_\alpha$. This is achieved by choosing $\hat \rho_I$ to be thermal equilibrium states of initial Hamiltonians $\hat{H}_I$, such that $[\hat{H}_I,\hat{H}_\alpha]\neq0$ and $[\hat{H}_I,\hat{H}_0]\neq0$. Dynamics are generated by quantum quenches in which, at time $\tau=0$, one suddenly changes $\hat{H}_I\rightarrow\hat{H}_\alpha$ and lets the system evolve unitarily. We consider systems that are translationally invariant before and after the quench. 

The time-evolving density matrix after the quench can be written as $\hat\rho(\tau)=e^{-i\hat{H}_\alpha\tau}\hat{\rho}_Ie^{i\hat{H}_\alpha\tau}$. We are interested in the dynamics of observables $\hat{O}$, whose expectation values are given by $O(\tau) = \text{Tr}\left[\hat{\rho}(\tau)\hat{O}\right]$. At long times, one expects observables to equilibrate at the values predicted by the diagonal ensemble (DE), $O_{\text{DE}}=\text{Tr}[\hat{\rho}_{\text{DE}}\hat{O}]$, where $\hat{\rho}_{\text{DE}} = \text{lim}_{\tau'\rightarrow\infty} (1/\tau') \int_0^{\tau'} \text{d}\tau\,\hat{\rho}(\tau)$ is the density matrix  of the diagonal ensemble \cite{rigol2008thermalization}. When written in the basis of eigenkets $\ket{E^\alpha_i}$ of $\hat{H}_\alpha$, $\hat{\rho}_{\text{DE}}$ takes the form
\begin{equation}
\hat{\rho}_{\text{DE}}=\sum_{i}\left(\bra{E^\alpha_i}\hat{\rho}_I\ket{E^\alpha_i}\right)\ket{E^\alpha_i}\bra{E^\alpha_i}\label{rho_DE} .
\end{equation}

For nonintegrable (quantum chaotic) systems, because of eigenstate thermalization~\cite{deutsch1991quantum, srednicki1994chaos, rigol2008thermalization}, one expects the predictions of the diagonal ensemble to match those of traditional statistical mechanics ensembles; namely, we expect observables to {\it thermalize}~\cite{dalessio_kafri_16}. For the $t$-$V$-$t'$-$V'$ Hamiltonian for hard-core bosons in 1D lattices, eigenstate thermalization and thermalization were studied in Ref.~\cite{rigol2009breakdown}, while quantum chaos was studied in Ref.~\cite{santos2010onset}. This means that, in our systems, we can also describe observables after equilibration by means of the grand canonical ensemble (GE) characterized by a temperature $T$, and, when particle number is a conserved quantity, by a chemical potential $\mu$. The grand canonical ensemble density matrices have the form
\begin{equation}\label{eq:rho_GE}
\hat{\rho}_{\text{GE}}=  
\begin{cases}
\dfrac{ e^{-\hat{H}_\alpha/T}}{ \text{Tr}[e^{-\hat{H}_\alpha/T}]} &  \text{when }g_\alpha\ne0\\ \ \\
\dfrac{ e^{-(\hat{H}_\alpha+\mu\hat{N})/T}}{ \text{Tr}[e^{-(\hat{H}_\alpha+\mu\hat{N})/T}]} & \text{when }g_\alpha=0.
\end{cases}
\end{equation}
  
When $g_\alpha\ne0$, $T$ is fixed by the (conserved) energy of the time-evolving state,
\begin{equation}
\text{Tr}[\hat\rho_\text{GE}\hat H_\alpha]=\text{Tr}[\hat \rho_{I}\hat H_\alpha]\label{finalT}.
\end{equation}
When $g_\alpha=0$, $T$ and $\mu$ are determined by the (conserved) energy [Eq.~\eqref{finalT}] and by the (conserved) particle number in the time-evolving state,
\begin{equation}
\text{Tr}[\hat\rho_\text{GE}\hat N]=\text{Tr}[\hat \rho_{I}\hat N]\label{finalmu}.
\end{equation}

Because of particle-hole symmetry in the Hamiltonians $\hat{H}_\alpha$, when $g_\alpha\neq0$ (namely, in the absence of particle-number conservation), our systems after equilibration are always at half filling irrespective of the initial filling $n_I$. We consider initial fillings $n_I\neq1/2$, which means that the filling must change during the dynamics when $g_\alpha\ne0$.

From our analysis in the previous sections, we expect that, for small values of $g_\alpha$, the dynamics of generic observables follow a two-step process towards thermalization: (i) fast relaxation driven by $\hat{H}_0$ (prethermal dynamics) and (ii) slower, nearly exponential, relaxation to the thermal equilibrium predictions (thermalization dynamics). At long times, close to thermal equilibrium, hydrodynamics may become dominant and algebraic relaxation is expected to take place~\cite{lux2014hydrodynamic}. That regime is not resolved in this work. 

As discussed in Sec.~\ref{sec:generalobservables}, the near-exponential dynamics following prethermalization can be described by projecting $\hat{\rho}(\tau)$ in the basis of the eigenkets of $\hat{H}_0$, up to an $\mathcal{O}(g)$ correction. This projected state is a diagonal ensemble of $\hat{H}_0$, whose density matrix [see Eq.~\eqref{eq:caPrho}] we denote as
\begin{equation}
\hat{\rho}_{0}(\tau)\equiv \caP_D \hat \rho(\tau) =\sum_{i}\left(\bra{E^0_i}\hat{\rho}(\tau)\ket{E^0_i}\right)\ket{E^0_i}\bra{E^0_i},\label{rho_0}
\end{equation}
where $\ket{E^0_i}$ are the eigenkets of $\hat{H}_0$.

\subsection{Numerical linked cluster expansion (NLCE)}\label{sec:nlce}

In what follows, we use the numerical linked cluster expansion (NLCE) approach introduced in Ref.~\cite{mallayya2018quantum} (see Refs.~\cite{whiteNLCE, guardado2017probing} for NLCE studies in two dimensions) to study the quantum dynamics of various observables in our translationally invariant 1D systems in the thermodynamic limit.

NLCEs were originally introduced to study systems in thermal equilibrium~\cite{rigol2006numerical, *rigol2007numerical1}, and allow one to obtain the expectation value of extensive observables per site (${\mathscr O}=O/L$), in the thermodynamic limit  ($L\rightarrow\infty$), as sums over the contributions of the connected clusters that can be embedded in the lattice. Given the connected clusters $c$, which can be embedded in the lattice in $M(c)$ ways per site and have weights $ W_{O}(c)$, one obtains ${\mathscr O}$ in the following way
\begin{equation}\label{nlce_eq}
\mathscr O=\sum_{c}M(c)\times W_{O}(c).
\end{equation}
$W_O(c)$ is computed, for each cluster $c$, from the expectation value of the observable $\hat{O}$ in the cluster ($O^c$) using the inclusion exclusion principle:
\begin{equation}\label{weight_subtraction}
W_{O}(c)=O^c- \sum_{s \subset c} W_{O}(s),
\end{equation}
where the sum is over all connected subclusters of $c$. For the smallest cluster, $W_{O}(c)=O^c$. For each cluster, $O^c=\text{Tr}[\hat{\rho}^c\hat{O}]$, where $\hat{\rho}^c$ is the density matrix of the relevant ensemble in the cluster. In NLCEs, $O^c$ is calculated exactly numerically using full exact diagonalization.

In our calculations, the density matrix of the initial state in each cluster $\hat{\rho}_I^c$ is taken to be the grand canonical density matrix set by the initial Hamiltonian in each cluster $\hat{H}_I^c$ (our initial states are in thermal equilibrium with respect to $\hat{H}_I$). In all quenches, we take $\hat{H}_I$ to be the $t$-$t'$-$V$-$V'$ model in Eq.~\eqref{model_H}. Since $[\hat{H}_I^c,\hat{N}^c]=0$, where $\hat{N}^c$ is the total particle number operator in the cluster, 
\begin{equation}
\hat{\rho}_I^c =\dfrac{ e^{-(\hat{H}_I^c+\mu_I\hat{N}^c)/T_I}}{\text{Tr}[e^{-(\hat{H}_I^c+\mu_I\hat{N}^c)/T_I}]}.
\end{equation}
Also, since $\hat{H}_I^c$ is particle-hole symmetric, we need $\mu_I\neq0$ in order to have initial states with filling $n_I\neq1/2$.

To calculate the time evolution of the expectation values of the observables ${\mathscr O}(\tau)$, where $\tau$ denotes the time after the quench, the density matrix of each cluster is evolved with the Hamiltonian after the quench $\hat{H}_\alpha^c$,
\begin{equation}
\hat{\rho}^c(\tau) = \left(e^{-i\hat{H}_\alpha^c\tau}\right)\hat{\rho}_I^c\left( e^{i\hat{H}_\alpha^c\tau}\right),
\end{equation}
and the NLCE calculation is carried out as usual~\cite{mallayya2018quantum}. Our Hamiltonians after the quench have the form in Eq.~\eqref{H_f}. Similarly, in order to obtain NLCE results after the quench for the diagonal ensemble, the grand canonical ensemble, and in the projected basis of $\hat{H}_0$, we use $\hat{\rho}^c_{\text{DE}}$, $\hat{\rho}^c_{\text{GE}}$, and $\hat{\rho}^c_{0}(\tau)$ from  Eqs.~\eqref{rho_DE}, \eqref{eq:rho_GE}, and \eqref{rho_0}, respectively, for each cluster $c$. 

NLCEs have been used to study quenches in the $t$-$t'$-$V$-$V'$ model [Eq.~\eqref{model_H}] to understand the dynamics~\cite{mallayya2018quantum}, and the description of observables after equilibration~\cite{rigol2014quantum, *rigol2016fundamental}, at the integrable point $t'=V'=0$ and away from it $t'=V'\neq0$. The presence of next-nearest neighbor hoppings and interactions makes it possible to have different building blocks to construct the clusters in the NLCE. In Ref.~\cite{mallayya2017numerical}, it was shown that maximally connected clusters---built adding contiguous sites and all possible bonds, starting from one site---are optimal for studying quenches in this model. Here, as in Refs.~\cite{mallayya2018quantum, rigol2014quantum, *rigol2016fundamental}, we use that NLCE in our calculations (there is only one maximally connected cluster with a given number of sites). The number of sites in the largest cluster considered defines the order of NLCE, and we denote the value of an observable $O(\tau)$ evaluated with NLCE to order $l$ as $O_l(\tau)$. 

\subsection{Observables}
We study three observables which have properties that make them qualitatively distinct in the context of dynamics and description after equilibration.

The first observable is the total particle number,
\begin{equation}
N(\tau)=\text{Tr}[\hat{\rho}(\tau)\hat{N}],
\end{equation}
whose value per site, the particle filling, is denoted as $n(\tau)$. This is a conserved quantity with respect to the reference Hamiltonian $\hat{H}_0$; i.e., it only changes during dynamics under $\hat{H}_\alpha$ after the quench if $g_\alpha\ne0$.

The second observable is the one-body nearest neighbor correlation,
\begin{equation}
K(\tau)=\text{Tr}[\hat{\rho}(\tau)\hat{K}],
\end{equation}
whose value per site is denoted as $k(\tau)$, where 
\begin{equation}
\hat{K}=\sum_i\left(\hat{b}_i^{\dagger}\hat{b}_{i+1}^{ }+\hat{b}_{i+1}^{\dagger}\hat{b}_i^{}\right).\label{KE_1}
\end{equation}
$k(\tau)$ is a local observable, closely related to the kinetic energy per site. It changes during the dynamics independently of whether $\hat{N}$ is conserved or not.

The third observable is the distribution function,  
\begin{equation}\label{eq:Mk}
M_k(\tau)=\text{Tr}[\hat{\rho}(\tau)\hat{M}_k],
\end{equation}
whose value per site is the momentum distribution function, denoted as $m_{k}(\tau)$. $\hat{M}_k$ is the unnormalized (to make each $k$ component extensive) Fourier transform of the one-body density matrix, given by
\begin{equation}\label{eq:fourierMk}
\hat M_{k}=\sum_{j,j'} e^{ik(j-j')} \hat b_j^{\dagger}\hat b_{j'}^{}.
\end{equation}   
$m_{k}(\tau)$ is a nonlocal one-body observable that changes during the dynamics independently of whether $\hat{N}$ is conserved or not. This observable is of particular interest because it is regularly measured (using time-of-flight expansion) in experiments with ultracold quantum gases~\cite{bloch2008many}. $m_{k}(\tau)$ was the observable used in Ref.~\cite{kinoshita2006quantum} to show lack of thermalization in 1D Bose gases with contact interactions, and in Ref.~\cite{tang_kao_18} to study prethermalization and thermalization 1D Bose gases with dipolar interactions.

\subsection{Parameters used in the calculations}\label{sec:calcparameters}

The initial state is taken to be a thermal equilibrium state at temperature $T_I=10$ and chemical potential $\mu_I=2$ (similar results are obtained for other values of $T_I$, not too low, and $\mu_I$), for an initial Hamiltonian $\hat{H}_I$ with nearest and next-nearest neighbor coupling parameters $t_I=0.5$, $V_I=1.5$, and $t'_I=V'_I=0.7$. After the quench, $\hat{H}_\alpha$ has coupling parameters $t=V=1$ (these set the energy scale in our calculations), $t'=V'=0.7$, and $g_\alpha\in(0, 0.12)$. For these parameters after the quench, $\hat{H}_0$~\cite{rigol2009breakdown, santos2010onset} and $\hat{H}_\alpha$ are quantum chaotic and the system thermalizes for all values of $g_\alpha$ (see Ref.~\cite{mallayya2018quantum} for a NLCE study of the quench dynamics when $g_\alpha=0$).

\begin{figure*}[!t]
\includegraphics[width=0.97\textwidth]{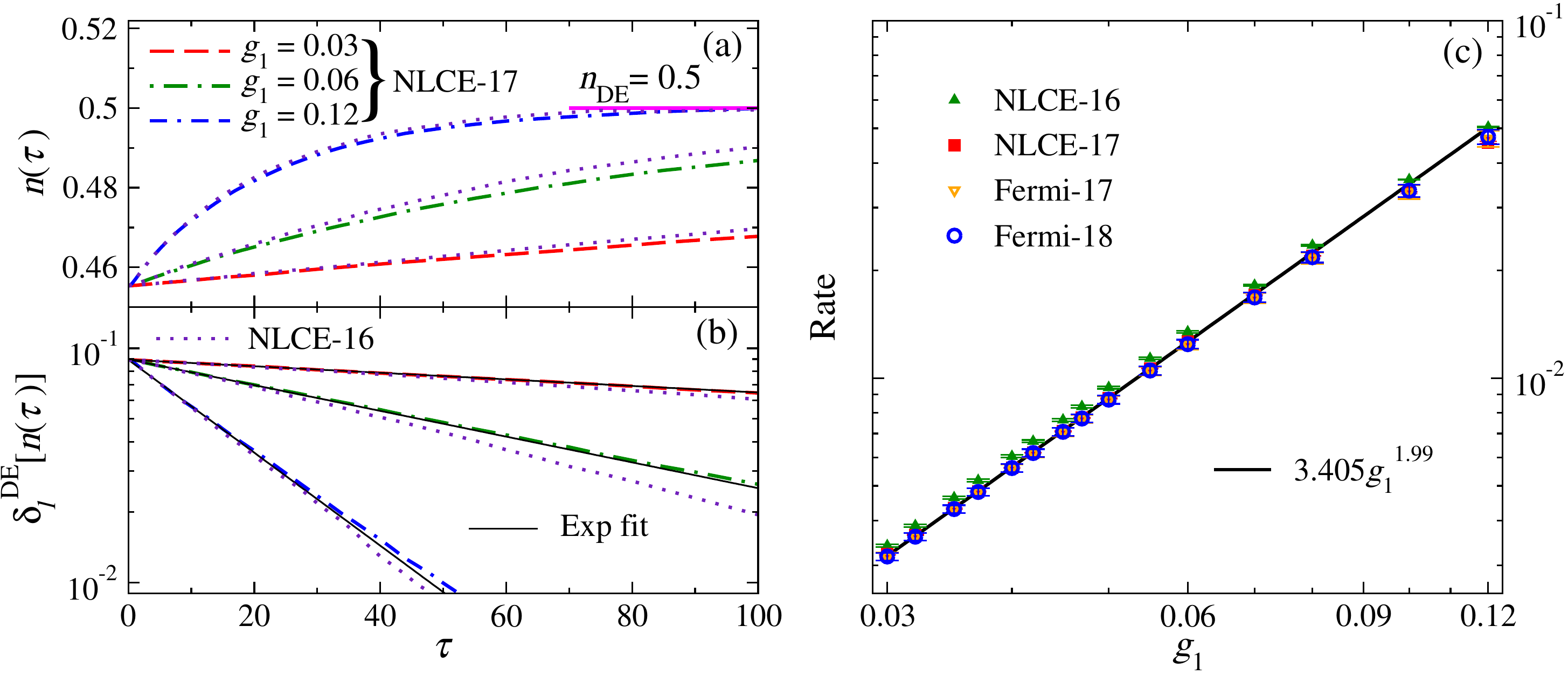}
\vspace{-0.1cm}
\caption{Dynamics of the particle filling after quantum quenches $\hat{H}_I\rightarrow\hat{H}_1$. (a) Dynamics of the particle filling $n(\tau)$ and (b) dynamics of the ``distance'' to equilibrium $\delta_l^{\text{DE}}\left[n(\tau)\right]$; see Eq.~\eqref{delta_n}. NLCE results are shown for $l=17$ [NLCE-17, see legend in (a)] and $l=16$ (NLCE-16, dotted lines). Straight lines in (b) depict fits to the results for $l=17$ in the interval $\tau\in[1,16]$ and to exponential functions $\propto\exp[-\Gamma^\text{NLCE}_{17}(g_1)\,\tau]$. (c) Thermalization rates $\Gamma^\text{NLCE}(g_1)$ (filled symbols) obtained from fits as the ones in (b), for $l=17$ in the interval $\tau\in[1,16]$ (NLCE-17), and for $l=16$ in the interval $\tau\in[1,6]$ (NLCE-16), reported for $g_1\in[0.03,0.12]$. Error bars show 95\% confidence bounds for the fits. The straight line is the result of a fit to $\Gamma^\text{NLCE}_{17}(g_1)\propto g_1^{\beta}$ for $g_1\in[0.03,0.06]$. The open symbols show the rates $\Gamma^\text{Fermi}(g_1)$ obtained evaluating Fermi's golden rule [see Eqs.~\eqref{fermi_rate} and~\eqref{fermi_gamma}] using full exact diagonalization in chains with $L=18$ (Fermi-18) and $L=17$ (Fermi-17), and periodic boundary conditions. The error bars show the standard deviation from averages over different choices of $\Delta E$ and $\tau$ (see Appendix~\ref{fermi_app}).}\label{part_num_fig}
\end{figure*}

For $n(\tau)$ and $k(\tau)$, we carry out the NLCE up to the 17th order for quenches with $g_\alpha\ne0$ (the largest Hamiltonian sector that needs to be diagonalized has 65\,792 states after exploiting reflection symmetry), and, thanks to particle number conservation, up to the 19th order for quenches with $g_\alpha=0$ (the dimension of the largest Hamiltonian sector that needs to be diagonalized is 46\,252). For $m_{k}(\tau)$, the NLCE is carried out to one order lower than for $n(\tau)$ and $k(\tau)$, namely, up to the 16th order for $g_\alpha\ne0$ and up to the 18th order for $g_\alpha=0$. This is because of the overhead generated by the calculation of the dynamics of all the matrix elements of the one-body density matrix [see Eqs.~\eqref{eq:Mk} and~\eqref{eq:fourierMk}]. 

\section{Numerical results}\label{sec:num}

\subsection{Dynamics of the particle filling}\label{sec:partfill}

In Fig.~\ref{part_num_fig}(a), we show the evolution of the particle filling under $\hat{H}_1$ for three values of $g_1$. Results are shown for the last two orders of the NLCE up to $\tau=100$. In those quenches, we expect the convergence errors for $n(\tau)$ to be below $0.01\%$ for times $\tau\lesssim 4$, and to remain low (below $1\%$) up to times $\tau\approx 16$ (see Appendix~\ref{NLCEvsED}). For $\tau\gtrsim 16$, the results for the last two orders of the NLCE can be seen to (slightly) deviate from each other in Fig.~\ref{part_num_fig}(a). In all the plots in Fig.~\ref{part_num_fig}(a), $n(\tau)$ can be seen to approach $n_\text{DE}=1/2$. For $g_1=0.12$, $n(\tau=100)\approx n_\text{DE}$.

The approach of $n(\tau)$ to $n_\text{DE}=0.5$ is exponential. This is apparent in Fig.~\ref{part_num_fig}(b), where we plot the normalized ``distance'' to equilibrium:
\begin{equation}\label{delta_n}
\delta_l^{\text{DE}}\left[n(\tau)\right]=\left|\dfrac{n_l(\tau)-n_{\text{DE}}}{n_{\text{DE}}}\right| \,.
\end{equation}     
We fit $\delta_l^\text{DE}\left[n(\tau)\right]$ to an exponential function $\propto \exp[-\Gamma^\text{NLCE}_l(g_1)\,\tau]$ to obtain the thermalization rate $\Gamma^\text{NLCE}_l(g_1)$. In order to gain an understanding of the accuracy of the obtained rates, we carry out fits in the time interval $\tau\in[1,16]$ for $l=17$ [the corresponding fits are shown in Fig.~\ref{part_num_fig}(b) as thin continuous lines], and in the interval $\tau\in[1,6]$ for $l=16$. The rates obtained in those calculations are shown in Fig.~\ref{part_num_fig}(c), as NLCE-17 and NLCE-16, respectively. They agree with each other within the errors of the fits. This suggests that our calculation of $\Gamma^\text{NLCE}(g_1)$ is robust. A power-law fit to the rates obtained for $l=17$ is also shown in Fig.~\ref{part_num_fig}(c). We find that $\Gamma^\text{NLCE}_{17}(g_1)\propto g_1^\beta$ with $\beta=1.99$, in agreement with the analytical results in Sec.~\ref{sec:slowdynamics}.

In closing Sec.~\ref{sec:deformedgge}, we argued that the rate $\dot q(\tau)=0$ whenever the state $\hat\rho(\tau)$ commutes exactly with $\hat{Q}$. The rate also vanishes if $\hat\rho(\tau)$ is time-reversal invariant about $\tau$. Both conditions apply to our initial states $\hat{\rho}_I$. As a result, there is a narrow plateau in $n(\tau)$ for $\tau\leq1$. This plateau is best seen in Fig.~\ref{part_num_NLCE_ED}. This is why, to obtain the rates reported in Fig.~\ref{part_num_fig}(c), we fit $n(\tau)$ at times $\tau\geq1$.

Next, we show that the values obtained for $\Gamma^\text{NLCE}(g_1)$ are in agreement with the ones predicted by Fermi's golden rule. From Eq.~\eqref{eq:drate}, changing $\hat{Q}\rightarrow\hat{N}$ and $\hat{V}\rightarrow\hat{V}_\alpha$, one can write for $\dot{n}(\tau)=dn/d\tau'|_{\tau'=\tau}$:
\begin{eqnarray}
\dot{n}(\tau)&=&\dfrac{2\pi g_\alpha^2}{L}\sum_{i,j} \delta(E^0_j-E^0_i) \left(N_j-N_i\right)P^0_i(\tau)\nonumber\\&&\times|\bra{E^0_j}\hat{V}_\alpha\ket{E^0_i}|^2,\label{fermi_rate}
\end{eqnarray} 
where $\ket{E^0_{i}}$ are the eigenkets of $\hat{H}_0$ with energies $E^0_{i}$, $N_i=\bra{E^0_i}\hat{N}\ket{E^0_i}$, and we average over the diagonal ensemble distribution, $P^0_i(\tau)=\bra{E^0_i}\hat{\rho}(\tau)\ket{E^0_i}$. To evaluate Eq.~\eqref{fermi_rate} numerically, we replace $\sum_{j} \delta(E^0_j-E^0_i)$ by a sum over energies $E^0_j$ that lie within a small energy window $\left[E^0_i-\Delta E/2, E^0_i+\Delta E/2\right]$ (see Appendix~\ref{fermi_app}).

The thermalization rates are estimated by computing
\begin{equation}
\Gamma^\text{Fermi}(g_\alpha)=-\dfrac{\dot{n}(\tau)}{n(\tau)-n_{\text{DE}}}\label{fermi_gamma}.
\end{equation}
When $\left|n(\tau)-n_{\text{DE}}\right|\ll n_{\text{DE}}$, $\Gamma^\text{Fermi}(g_\alpha)$ becomes independent of $\tau$ and $n(\tau)$ relaxes exponentially. This condition is, to a good degree, satisfied for our choice of initial state, for which $\delta_l^{\text{DE}}\left[n(\tau)\right]< 0.09$ [see Fig.~\ref{part_num_fig}(b)]. Our calculations of $\Gamma^\text{Fermi}(g_\alpha)$ are done using full exact diagonalization in chains with $L=17$ and 18 sites, and periodic boundary conditions. We identify a range of values for $\Delta E$ and $\tau$ for which the results for $\Gamma^\text{Fermi}(g_\alpha)$ are robust against the choice of $\Delta E$, $\tau$, and $L$ (see Appendix~\ref{fermi_app}).

In Fig.~\ref{part_num_fig}(c), we report our results for $\Gamma^\text{Fermi}(g_1)$. They are in excellent agreement with $\Gamma^\text{NLCE}(g_1)$. We should add that, for quenches $\hat{H}_I\rightarrow\hat{H}_2$, Eq.~\eqref{fermi_rate} predicts the same leading $\caO(g_\alpha^2)$ dynamics as under $\hat{H}_1$. This is the case because the terms that change the total particles number are the same in $\hat{V}_1$ and $\hat{V}_2$. Hence, $n(\tau)$ is the same for both Hamiltonians up to corrections $\caO(g_\alpha^3)$.
 
\subsection{Dynamics of the one-body nearest neighbor correlation}\label{sec:nearhopp}

\begin{figure}[!b]
\includegraphics[width=0.83\columnwidth]{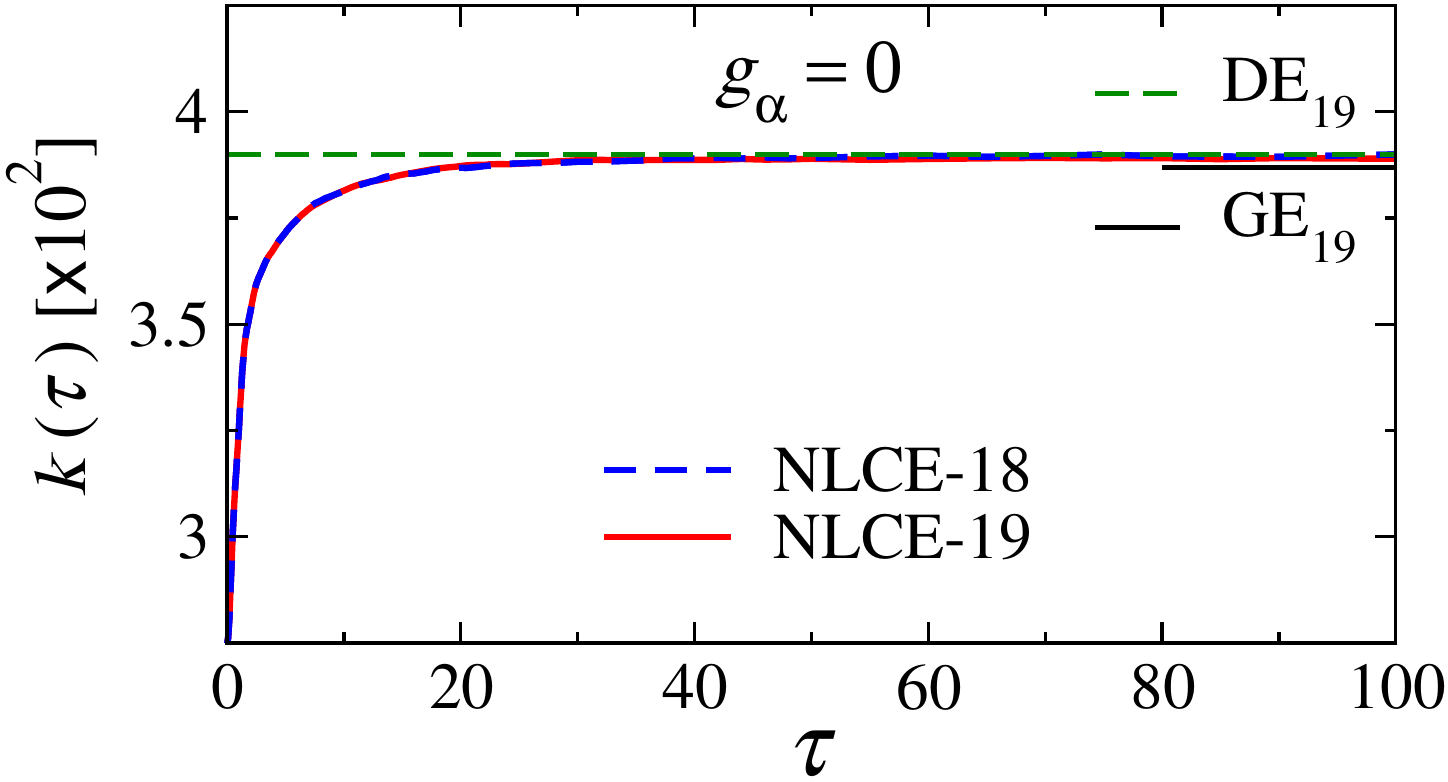}
\vspace{-0.1cm}
\caption{Dynamics of the one-body nearest neighbor correlation $k(\tau)$ when $g_\alpha=0$ after the quench. Results are shown for the 18th (NLCE-18) and 19th (NLCE-19) orders of the NLCE. We also report, as horizontal lines, the results for the 19th order of the diagonal ensemble (DE\ts{19}) and for the 19th order of the grand canonical ensemble (GE\ts{19}).}\label{hopping_unperturbed}
\end{figure}

Next, we study the dynamics of the one-body nearest neighbor correlation, $k(\tau)$ [see Eq.~\eqref{KE_1}]. In contrast to the particle filling studied in Sec.~\ref{sec:partfill}, the nearest neighbor correlation $k(\tau)$ exhibits dynamics even if $g_\alpha=0$ after the quench. 

Figure~\ref{hopping_unperturbed} shows the 18th (NLCE-18) and 19th (NLCE-19) orders of the NLCE for $k(\tau)$ after a quench in which $g_\alpha=0$. The results of both orders are in excellent agreement with each other, and equilibrate rapidly to the prediction of the diagonal ensemble (horizontal dashed line). The predictions of the diagonal and grand canonical (horizontal contiguous line) ensembles are very close to each other, indicating thermalization. This is expected because $\hat{H}_0$ after the quench is nonintegrable. The small difference between the diagonal and the grand canonical ensemble results is due to the lack of convergence of the NLCE for the former (the latter is fully converged)~\cite{rigol2014quantum, *rigol2016fundamental}. Those differences decrease with increasing the order of the NLCE. In what follows, we use the grand canonical ensemble predictions to probe thermalization.

\begin{figure}[!t]
\includegraphics[width=0.985\columnwidth]{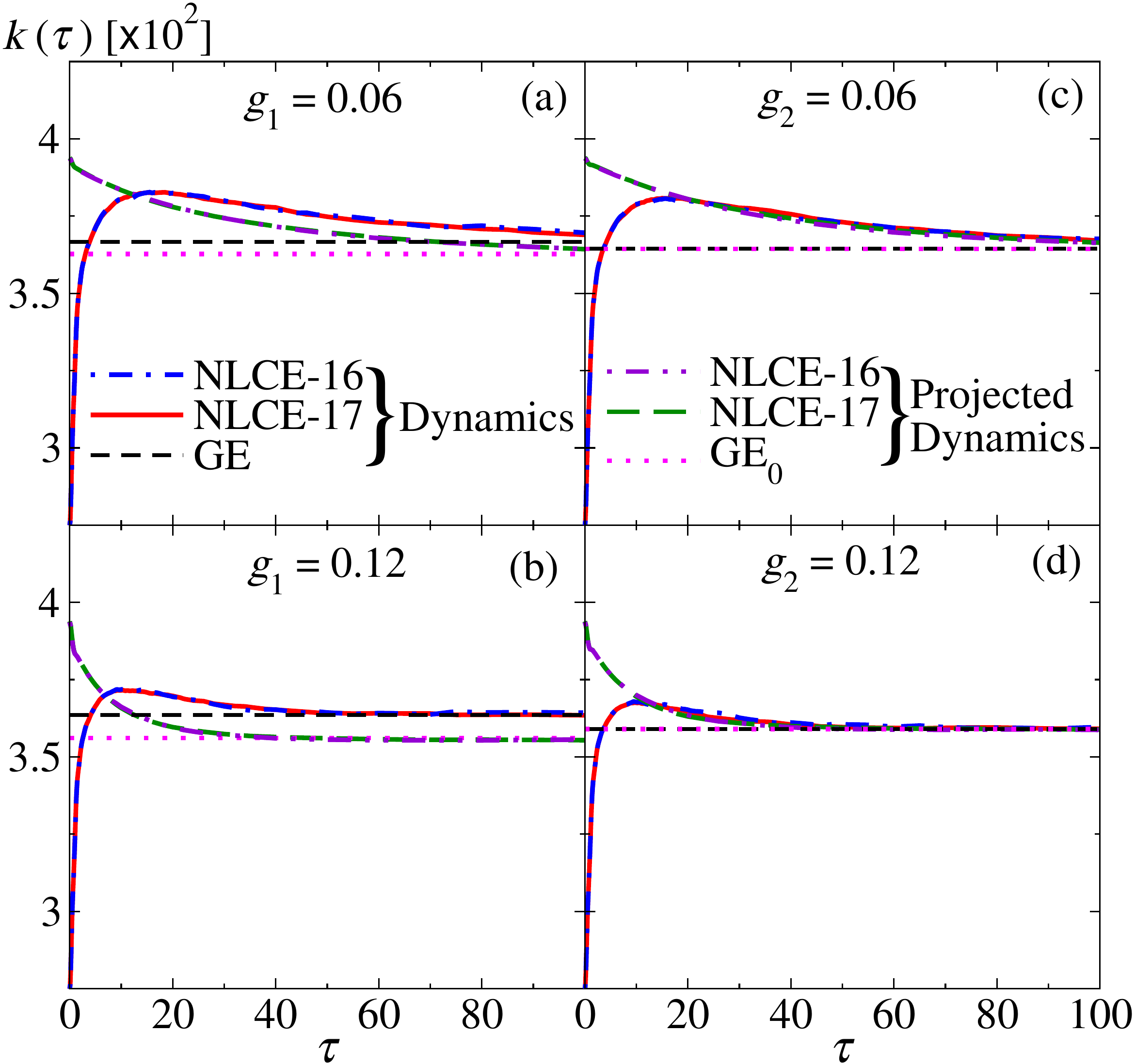}
\vspace{-0.1cm}
\caption{Dynamics of the one-body nearest neighbor correlation $k(\tau)$ for (a) $g_1=0.06$, (b) $g_1=0.12$, (c) $g_2=0.06$, and (d) $g_2=0.12$. Results are shown for the 16th (NLCE-16) and 17th (NLCE-17) orders of the NLCE, both for the dynamics [see legends in (a)] and for the dynamics in the projected basis of $\hat{H}_0$ [see legends in (c)]. The horizontal lines show the results for the grand canonical ensemble corresponding to the original dynamics (GE) and to the projected dynamics (GE$_0$), both evaluated at the 17th order of the NLCE.}\label{hopping_perturbed}
\end{figure}

In Fig.~\ref{hopping_perturbed}, we show $k(\tau)$ after quenches $\hat{H}_I\rightarrow\hat{H}_1$ (left-hand panels) and $\hat{H}_I\rightarrow\hat{H}_2$ (right-hand panels). The dynamics are qualitatively similar in both cases. They can be split in two regimes: (i) fast (prethermal) dynamics driven by $\hat{H}_0$ (note that, at times $\tau\lesssim 10$, dynamics in Fig.~\ref{hopping_perturbed} are nearly identical to those in Fig.~\ref{hopping_unperturbed}) and (ii) a slower (thermalization) dynamics controlled by the strength of the perturbation. During the latter regime, the system approaches [and reaches for $g_1=g_2=0.12$ and $\tau=100$, see Figs.~\ref{hopping_perturbed}(b) and~\ref{hopping_perturbed}(d)] the prediction of the grand canonical ensemble $\hat \rho_{\text{GE}}$ corresponding to $\hat{H}_\alpha$ after the quench (with the temperature set by the initial state).

The slow approach to thermal equilibrium can be well described using the projected state $\hat{\rho}_0(\tau)$ from Eq.~\eqref{rho_0}. In Fig.~\ref{hopping_perturbed}, we show the results for the projected dynamics along with those for the actual dynamics. As follows from the discussion in Sec.~\ref{sec:generalobservables}, the results for the projected dynamics approach those of a thermal equilibrium state of $\hat{H}_0$, which we denote as $\hat \rho_{\text{GE}_0}$. We compute the temperature $T_0$ in $\hat \rho_{\text{GE}_0}$ using the expectation value of $\hat{H}_0$ in the thermal equilibrium state $\hat \rho_{\text{GE}}$ of $\hat{H}_\alpha$:
\begin{equation}
\text{Tr}[\hat\rho_{\text{GE}_0}\hat{H}_0]=\text{Tr}[\hat \rho_{\text{GE}}\hat H_0]\label{finalT_0}.   
\end{equation}
The chemical potential in $\hat\rho_{\text{GE}_0}$ is $\mu_0=0$ because, for $g_\alpha\neq0$, the systems after equilibration are at half filling.

\begin{figure}[!t]
\includegraphics[width=0.9\columnwidth]{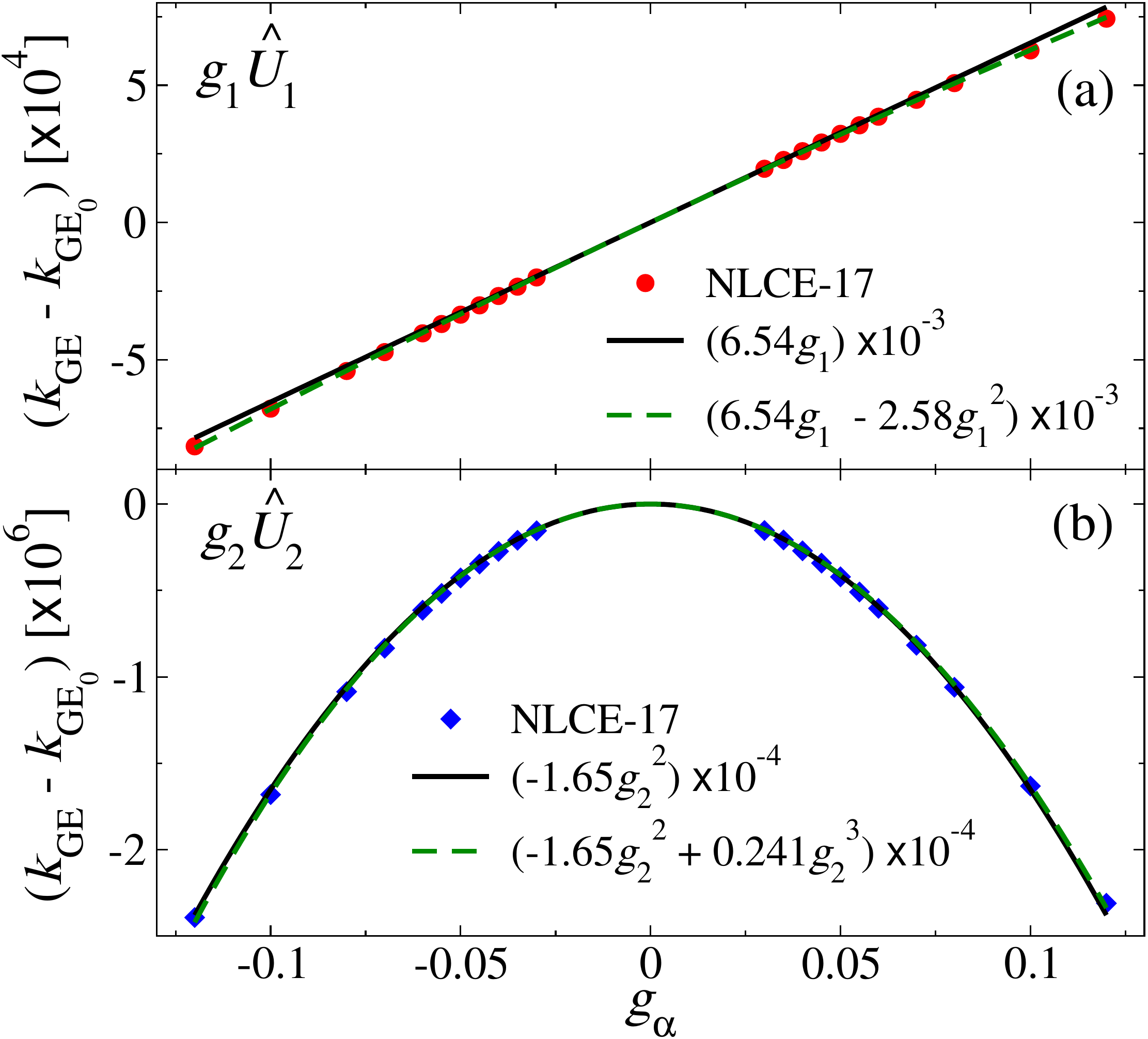}
\vspace{-0.1cm}
\caption{Difference in the equilibrium (grand canonical ensemble) results between the actual dynamics (GE) and the projected dynamics (GE$_0$) for the one-body nearest neighbor correlation ($k_{\text{GE}}-k_{\text{GE}_0})$ evaluated at the 17th order of the NLCE. (a) $g_1\hat{V}_1$ perturbation (symbols), with linear and linear plus quadratic fits (lines). (b) $g_2\hat{V}_2$ perturbation (symbols), with quadratic and quadratic plus cubic fits (lines).}\label{nhop_correction}
\end{figure}

In the left-hand panels in Fig.~\ref{hopping_perturbed}, for quenches $\hat{H}_I\rightarrow\hat{H}_1$, one can see the advanced offsets (see Sec.~\ref{sec:generalobservables}) between the dynamics and the projected dynamics. The offsets are much smaller for quenches $\hat{H}_I\rightarrow\hat{H}_2$, whose results are shown in the right-hand panels in Fig.~\ref{hopping_perturbed}. The offsets between the actual and projected dynamics remain constant at long times and are, essentially, the difference between the predictions of $\hat \rho_{\text{GE}}$ and $\hat\rho_{\text{GE}_0}$. 

The difference in the offsets generated by quenches $\hat{H}_I\rightarrow\hat{H}_1$ and $\hat{H}_I\rightarrow\hat{H}_2$ can be understood using Eq.~\eqref{eq:correctiontoplateau}, replacing $\hat{O}\rightarrow\hat{K}$ [$\hat{K}$ is defined in Eq.~\eqref{KE_1}] and $\hat{V}\rightarrow\hat{V}_\alpha$. Since $\hat{K}$ and $\hat{\rho}_0(\tau)$ [$\hat{\rho}_0(\tau)$ is defined in Eq.~\eqref{rho_0}] are block diagonal in the particle number basis, only the presence of terms in $\hat{V}_\alpha$ that do not change the particle number can produce an $\mathcal{O}(g)$ correction. Such terms are present in $\hat{V}_1$ (the hopping terms), see Eq.~\eqref{V1}, but are absent in $\hat{V}_2$, see Eq.~\eqref{V2}. This means that, to leading order, $\Delta k_1(\tau)\propto g_1$ while $\Delta k_2(\tau)\propto g_2^\beta$ with $\beta\geq2$. 

In Fig.~\ref{nhop_correction}, we show the long-time offsets between the actual and projected dynamics as functions of $g_\alpha$ (for positive and negative values of $g_\alpha$) in the quenches $\hat{H}_I\rightarrow\hat{H}_1$ [Fig.~\ref{nhop_correction}(a)] and $\hat{H}_I\rightarrow\hat{H}_2$ [Fig.~\ref{nhop_correction}(b)]. The offsets are computed as the difference between the predictions of $\hat \rho_{\text{GE}}$ and $\hat\rho_{\text{GE}_0}$ for the equilibrated results (see the horizontal lines in Fig.~\ref{hopping_perturbed}). Those predictions are converged to machine precision in the 17th order of the NLCE shown in Fig.~\ref{nhop_correction}. Figure~\ref{nhop_correction}(a), for quenches $\hat{H}_I\rightarrow\hat{H}_1$, makes apparent the presence of a leading linear correction and of a subleading quadratic one. Figure~\ref{nhop_correction}(b) shows the absence of the linear correction for quenches $\hat{H}_I\rightarrow\hat{H}_2$. There, the leading correction is quadratic, and our numerical results allow us to identify a subleading cubic correction, which leads to a weak asymmetry about $g_2=0$.

Comparing the results reported in Fig.~\ref{part_num_fig}(a), and in Figs.~\ref{hopping_perturbed}(a) and~\ref{hopping_perturbed}(b), for $g_1=0.06$ and 0.12, it becomes apparent that $k(\tau)$ equilibrates (reaches the long-time horizontal line prediction) faster than $n(\tau)$. This can be understood to be the result of the one-body nearest neighbor correlation being particle-hole symmetric (as the Hamiltonians $\hat{H}_\alpha$ are). This means that, close to equilibrium, the difference between $k(\tau)$ and $k_\text{GE}$ can only be a function of even powers of the difference between $n(\tau)$ and $n=1/2$. Say the leading even power in the difference between $n(\tau)$ and $n=1/2$ entering in $k(\tau)-k_\text{GE}$ is $2$, then $\dot{k}(\tau)\simeq 2 \Gamma^\text{Fermi}(g_\alpha)\, [k(\tau)-k_\text{GE}]$. Our numerical results for the thermalization rates of $k(\tau)$, not shown, support the correctness of this simple analysis.

\begin{figure}[!b]
\includegraphics[width=0.985\columnwidth]{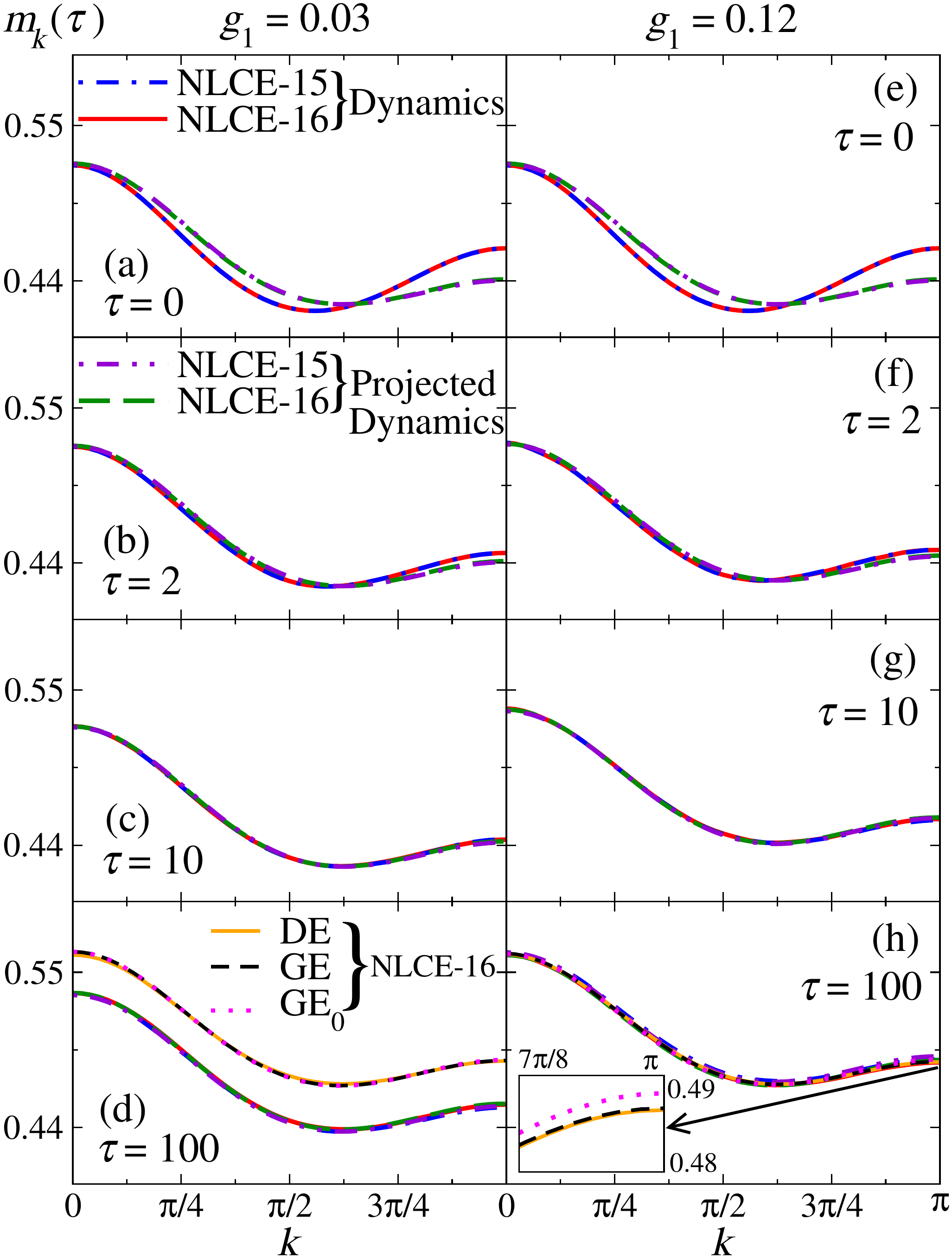}
\vspace{-0.2cm}
\caption{Momentum distribution function $m_k(\tau)$ [see Eq.~\eqref{eq:Mk}] at four times after quenches in which $\hat{H}_I\rightarrow\hat{H}_1$ with $g_1=0.03$ [(a)--(d)] and $g_1=0.12$ [(e)--(h)]. $m_k(\tau)$ is evaluated at $\tau=0$ [(a),(e)], $\tau=2$ [(b),(f)], $\tau=10$ [(c),(g)], and $\tau=100$ [(d),(h)]. Results are reported for the 15th (NLCE-15) and 16th (NLCE-16) orders of the NLCE. We show $m_k(\tau)$ for the dynamics [see the legends in (a)], and for the projected dynamics in the basis of $\hat{H}_0$ [see the legends in (b)]. The predictions of the diagonal ensemble (DE) and the grand canonical ensemble (GE) for the dynamics, and of the grand canonical ensemble for the projected dynamics (GE\ts0), all evaluated at the 16th order of the NLCE, are shown in (d) and (h) [see the legends in (d)]. The inset in (h) shows the predictions of the latter three ensembles about $k=\pi$.}\label{mom_dist}
\end{figure}

\begin{figure*}[!t]
\includegraphics[width=0.97\textwidth]{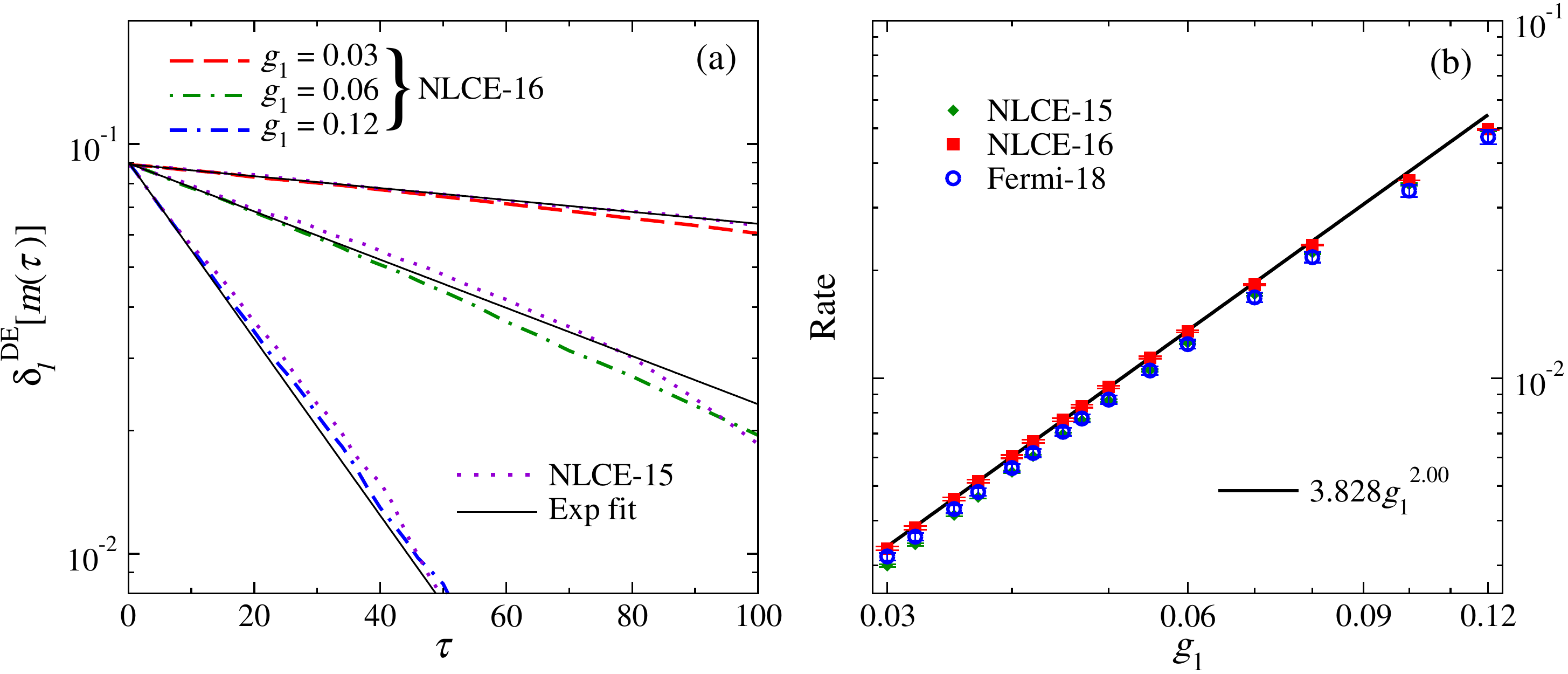}
\vspace{-0.1cm}
\caption{Dynamics of the momentum distribution function $m_k(\tau)$ [Eq.~\eqref{eq:Mk}] after quenches $\hat{H}_I\rightarrow\hat{H}_1$. (a) Dynamics of the ``distance'' to equilibrium $\delta^{\text{DE}}\left[m(\tau)\right]$; see Eq.~\eqref{delta_mk}. NLCE results are shown for $l=16$ [NLCE-16] and $l=15$ (NLCE-15, dotted lines). Straight lines in (a) depict fits to the results for $l=16$ in the interval $\tau\in[2,6]$ and to exponential functions $\propto\exp[-{\it\Gamma}^\text{NLCE}_{16}(g_1)\,\tau]$. (b) Thermalization rates ${\it\Gamma}^\text{NLCE}(g_1)$ of $m_k(\tau)$ (filled symbols) obtained from fits as the ones in (a), for $l=16$ in the interval $\tau\in[2,6]$ (NLCE-16), and for $l=15$ in the interval $\tau\in[2,5]$ (NLCE-16), reported for $g_1\in[0.03,0.12]$. Error bars show 95\% confidence bounds for the fits. The straight line is the result of a fit to ${\it\Gamma}^\text{NLCE}_{16}(g_1)\propto g_1^{\beta}$ for $g_1\in[0.03,0.06]$. The open circles show the rates $\Gamma^\text{Fermi}(g_1)$ [also reported in Fig.~\ref{part_num_fig}(c)] obtained by evaluating Fermi's golden rule [see Eqs.~\eqref{fermi_rate} and~\eqref{fermi_gamma}] using full exact diagonalization in chains with $L=18$ (Fermi-18) and periodic boundary conditions. The error bars show the standard deviation from averages over different choices of $\Delta E$ and $\tau$ (see Appendix~\ref{fermi_app}).}\label{mom_rate}
\end{figure*}

\subsection{Dynamics of the momentum distribution}\label{sec:momentum}

Here, we study the dynamics of the momentum distribution function $m_k(\tau)$ [Eq.~\eqref{eq:Mk}] for quenches $\hat{H}_I\rightarrow\hat{H}_1$. Similar to the dynamics of $k(\tau)$, the evolution of $m_k(\tau)$ can be split into a fast prethermalization dynamics and a slower relaxation to the thermal equilibrium prediction at a rate controlled by the strength of the perturbation.

In Fig.~\ref{mom_dist}, we show results for the 15th (NLCE-15) and 16th (NLCE-16) orders of the NLCE for $m_k(\tau)$ at four times ($\tau=0$, 2, 10, and 100) after quenches in which $g_1=0.03$ (left-hand panels in Fig.~\ref{mom_dist}) and $g_1=0.12$ (right-hand panels in Fig.~\ref{mom_dist}). In all panels in Fig.~\ref{mom_dist}, we also show results for the 15th (NLCE-15) and 16th (NLCE-16) orders of the NLCE for the projected dynamics of $m_k(\tau)$ in the basis of $\hat{H}_0$ [the dynamics dictated by $\hat{\rho}_0(\tau)$, see Eq.~\eqref{rho_0}]. Figures~\ref{mom_dist}(b) and~\ref{mom_dist}(f) show that the momentum distribution in the original and projected dynamics become nearly indistinguishable from each other after short times ($\tau\gtrsim2$), and remain so at long times [see Figs.~\ref{mom_dist}(c),~\ref{mom_dist}(d),~\ref{mom_dist}(g), and~\ref{mom_dist}(h)]. 

In Figs.~\ref{mom_dist}(d) and~\ref{mom_dist}(h), we also show the 16th order NLCE prediction of the diagonal and the grand canonical ensembles for the long-time dynamics, and of the grand canonical ensemble for the long-time projected dynamics [see Eq.~\eqref{finalT_0} and the discussion surrounding this equation]. For $g_1=0.03$ [Fig.~\ref{mom_dist}(d)], the results of the three equilibrium ensembles are nearly indistinguishable from each other. They differ from those of the dynamics at $\tau=100$ (they all become nearly indistinguishable from each other at later times). For $g_1=0.12$ [Fig.~\ref{mom_dist}(h)], the results for the three equilibrium ensembles and for the dynamics at $\tau=100$ agree with each other. This, in contrast to the results for $g_1=0.03$, makes apparent that $m_k(\tau)$ thermalizes faster with increasing the magnitude of $g_1$ (as expected). Also, as expected from our discussion in Sec.~\ref{sec:generalobservables} and for $k(\tau)$ in Sec.~\ref{sec:nearhopp} (see Fig.~\ref{hopping_perturbed}), there is an $\mathcal{O}(g_1)$ offset between the results for $m_k(\tau)$ in the dynamics and in the projected dynamics. The magnitude of this offset is momentum dependent, as made apparent by the inset in Fig.~\ref{mom_dist}(d), and is too small to be resolved at the scales used in the main panels of Fig.~\ref{mom_dist}.
 
The operator corresponding to the momentum distribution function, unlike the one for the one-body nearest neighbor correlation, is not particle-hole symmetric. This implies that, as $m_k(\tau)$ approaches its equilibrium value in the diagonal ensemble $m_{k}^{\text{DE}}$, to leading order
\begin{eqnarray}
m_k(\tau)-m_{k}^{\text{DE}}\propto\left[n(\tau)-\dfrac{1}{2}\right].
\end{eqnarray}
Thus, we expect $m_k$ to thermalize with the same rate $\Gamma^{\text{Fermi}}(g_1)$ given by Eq.~\eqref{fermi_gamma}. To quantify the ``distance'' to equilibrium for $m_k(\tau)$, we compute
\begin{equation}\label{delta_mk}
\delta^{\text{DE}}_l\left[m(\tau)\right]=\dfrac{\sum_k\left|m^l_k(\tau)-m_{k}^{l,\text{DE}}\right|}{\sum_k m_{k}^{l,\text{DE}}}.
\end{equation} 

In Fig.~\ref{mom_rate}(a), we show $\delta^\text{DE}_l\left[m(\tau)\right]$ evaluated at the 15th (NLCE-15) and 16th (NLCE-16) orders of the NLCE. $\delta^\text{DE}_{15}\left[m(\tau)\right]$ and $\delta^\text{DE}_{16}\left[m(\tau)\right]$ can be seen to decay close to exponentially, although the convergence of the results for $\delta^\text{DE}_l\left[m(\tau)\right]$ in Fig.~\ref{mom_rate}(a) is not as good as for $\delta^\text{DE}_l\left[n(\tau)\right]$ in Fig.~\ref{part_num_fig}(b). This is understandable because (i) we are able to calculate one order lower for $m_k(\tau)$ than for $n(\tau)$ and (ii) $m_k(\tau)$ probes correlations at all distances, while $n(\tau)$ is local and is a thermodynamic quantity.

We fit $\delta_l^\text{DE}\left[m(\tau)\right]$ to an exponential function $\propto\exp[-{\it\Gamma}^\text{NLCE}_l(g_1)\,\tau]$ to obtain the thermalization rates ${\it\Gamma}^\text{NLCE}_l(g_1)$ for the momentum distribution function. The fits are carried out in the interval $\tau\in[2,6]$ for NLCE-16 [shown as thin continuous lines in Fig.~\ref{mom_rate}(a)], and in the interval $\tau\in[2,5]$ for NLCE-15. The rates ${\it\Gamma}^\text{NLCE}_l(g_1)$ are reported in Fig.~\ref{mom_rate}(b) for $g_1\in[0.03,0.12]$. In Fig.~\ref{mom_rate}(b), we also plot the rates $\Gamma^{\text{Fermi}}(g_1)$ obtained by evaluating Fermi's golden rule [see Eqs.~\eqref{fermi_rate} and~\eqref{fermi_gamma}] using full exact diagonalization in chains with $L=18$ sites and periodic boundary conditions (see Sec.~\ref{sec:partfill} and Appendix~\ref{fermi_app}). [The rates $\Gamma^{\text{Fermi}}(g_1)$ were also reported in Fig.~\ref{part_num_fig}(c).] Figure~\ref{mom_rate}(b) shows that, as advanced, the thermalization rates for the momentum distribution function are the same (within our computational errors) as the ones for the particle filling. A power-law fit to the rates ${\it\Gamma}^\text{NLCE}_{16}(g_1)$ is also shown in Fig.~\ref{mom_rate}(b). We find that ${\it\Gamma}^\text{NLCE}_{16}(g_1)\propto g_1^\beta$ with $\beta=2.00$, in agreement with the numerical results in Sec.~\ref{sec:partfill}, and with the analytical ones in Secs.~\ref{sec:slowdynamics} and~\ref{sec:generalobservables}.

\section{Summary and discussion}\label{sec:summary}

We put forward a conceptually simple scenario for prethermalization and thermalization in isolated quantum systems with a weakly broken conservation law. This scenario applies equally to noninteracting and strongly interacting integrable systems in which integrability is weakly broken, as well as to nonintegrable systems in which a conservation law is weakly broken. The weak perturbation allows the system to equilibrate to a state that is a (generalized) thermal equilibrium state of the unperturbed system. The properties of such a state are determined by the slowly changing value of the quasiconserved quantity (or quantities). The separation of timescales leads to a universal description, two aspects of which stand out.

\noindent(i) The dynamics of the (or each) quasiconserved quantity is described by an autonomous equation that can be constructed from Fermi's golden rule in unperturbed equilibrium ensembles. This equation is the generalization of the nonlinear Boltzmann equation appearing in weakly interacting quantum systems. 

\noindent(ii) The deviation of observables in the instantaneous state from the prediction of the unperturbed equilibrium ensemble is described by first-order perturbation theory. This generalizes the concept of ``deformed GGE'' that was described in Ref.~\cite{essler2014quench} for integrable systems in the presence of a weak integrability-breaking perturbation.

\begin{figure}[!t]
\includegraphics[width=0.985\columnwidth]{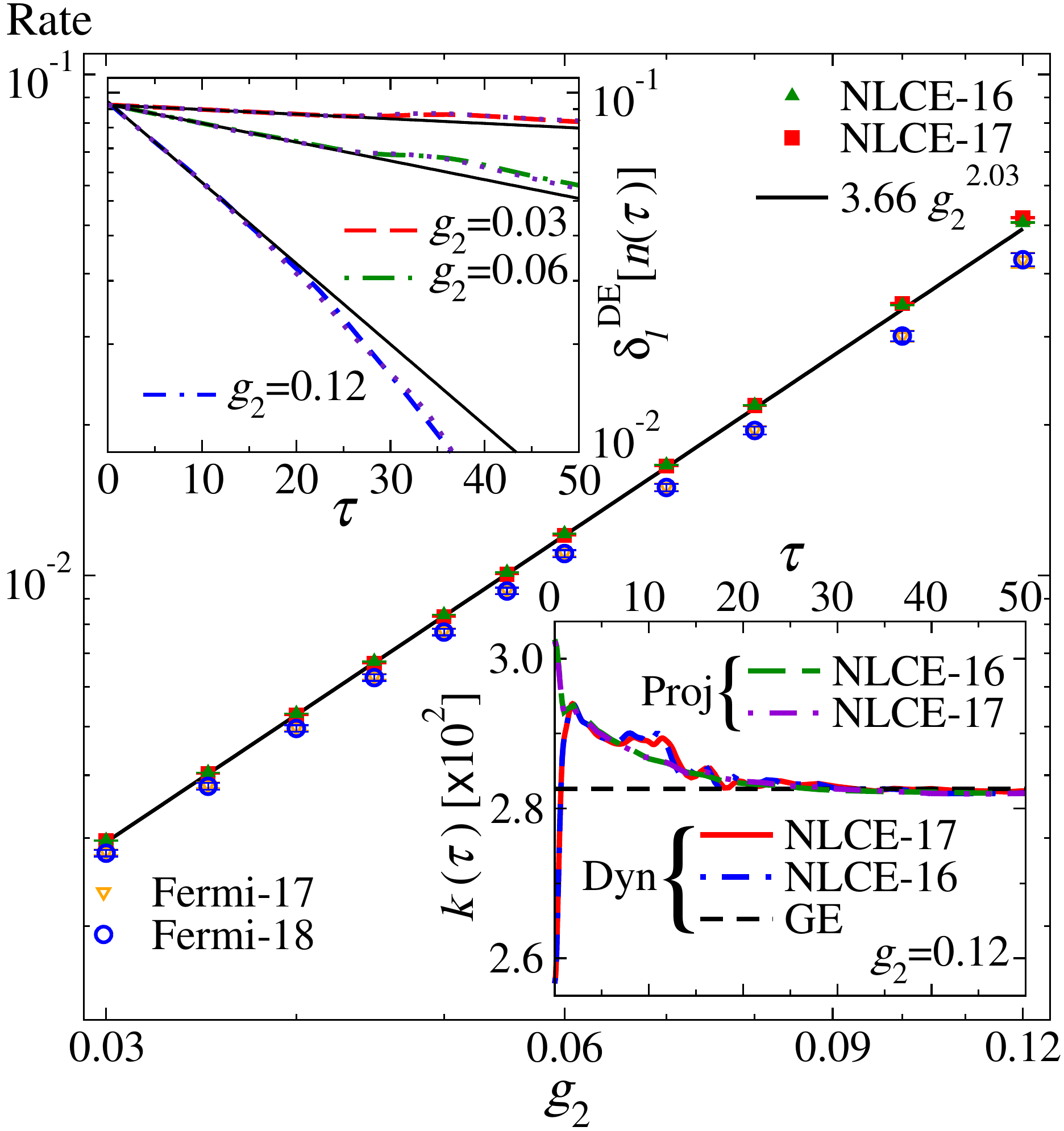}
\vspace{-0.1cm}
\caption{Dynamics after quantum quenches $\hat{H}_I\rightarrow\hat{H}_2$, for $t'_I=V'_I=0$ in $\hat{H}_I$ and $t'=V'=0$ in $\hat H_2$ (integrable reference dynamics). Main panel: Thermalization rates $\Gamma^\text{NLCE}(g_2)$ for the particle filling dynamics (filled symbols) obtained from fits as the ones shown in the upper inset. The straight line is the result of a fit to $\Gamma^\text{NLCE}_{17}(g_2)\propto g_2^{\beta}$. The open symbols show the rates $\Gamma^\text{Fermi}(g_2)$, see Eqs.~\eqref{fermi_rate} and~\eqref{fermi_gamma}, obtained using full exact diagonalization (see Appendix~\ref{fermi_app}). Upper inset: $\delta_l^{\text{DE}}\left[n(\tau)\right]$, see Eq.~\eqref{delta_n}, versus $\tau$. The straight lines depict fits to exponential functions $\propto\exp[-\Gamma^\text{NLCE}_{17}(g_2)\,\tau]$. The legends in the main panel and in the upper inset follow Fig.~\ref{part_num_fig}, and the fits are done in the same intervals as in Fig.~\ref{part_num_fig}. Lower inset: Dynamics, and projected dynamics in the basis of $\hat{H}_0$, of $k(\tau)$ for $g_2=0.12$. The horizontal line shows the result for the grand canonical ensemble corresponding to the original dynamics. The legends are identical to those in Fig.~\ref{hopping_perturbed}.}\label{part_num_integrable}
\end{figure}

Our theoretical results, as well as several special behaviors related to the initial state selected, properties of the perturbations that break the conservation law, and properties of the observables studied, were validated by numerical experiments in systems for which both the reference and the perturbed dynamics are nonintegrable.

A systematic study of integrable systems in which integrability is weakly broken is beyond the scope of this work. As mentioned in the Introduction, noninteracting systems in the presence of weak integrability-breaking interactions were studied in Refs.~\cite{moeckel_kehrein_2008, *moeckel_kehrein_2009, eckstein_kollar_09, kollar_wolf_11, tavora_mitra_13, nessi_iucci_14, essler2014quench, bertini2015prethermalization, *bertini2016prethermalization, fagotti2015universal, reimann_dabelow_19}. There have also been studies of integrable systems mappable onto noninteracting ones in the presence of an integrability-breaking perturbation~\cite{Marcuzzi_2013, *Marcuzzi_2016}. For a strongly interacting integrable system, $\hat{H_0}$ in Eq.~\eqref{model_H} with $t=V=1$ and $t'=V'=0$ (which is not mappable onto a noninteracting model), in Ref.~\cite{mallayya2018quantum} it was shown numerically that weakly breaking integrability by making $g\equiv t'=V'\neq0$ results in thermalization rates $\propto g^2$.

To check the applicability of our theory to strongly interacting integrable reference dynamics, in Fig.~\ref{part_num_integrable} we show results for dynamics under the same reference Hamiltonian as in Ref.~\cite{mallayya2018quantum} when one breaks integrability with the perturbation $g_2\hat{V}_2$ in Eq.~\eqref{V2}. Those results are the equivalent of results reported in Figs.~\ref{part_num_fig} and~\ref{hopping_perturbed}, for quenches $\hat{H}_I\rightarrow\hat{H}_2$ in which $t'_I=V'_I=0$ in $\hat H_I$ and $t'=V'=0$ in $\hat H_2$. The main panel in Fig.~\ref{part_num_integrable} shows that the thermalization rate for the particle filling is $\propto g_2^2$, and that it agrees with the Fermi golden rule prediction [see Eqs.~\eqref{fermi_rate} and~\eqref{fermi_gamma}] for small values of $g_2$. The upper inset in Fig.~\ref{part_num_integrable} shows that the particle filling approaches exponentially its thermal equilibrium value, as seen in Fig.~\ref{part_num_fig}(b) for the nonintegrable reference dynamics. The lower inset in Fig.~\ref{part_num_integrable} shows the dynamics, and the projected dynamics in the basis of $\hat{H}_0$, of $k(\tau)$ for $g_2=0.12$. They are in excellent agreement with each other, like in Fig.~\ref{hopping_perturbed}(d) for the nonintegrable reference dynamics. In short, there are no qualitative differences between the results reported in Figs.~\ref{part_num_fig} and~\ref{hopping_perturbed} for the nonintegrable reference dynamics and in Fig.~\ref{part_num_integrable} for the integrable reference dynamics.

\begin{figure}[!t]
\includegraphics[width=0.985\columnwidth]{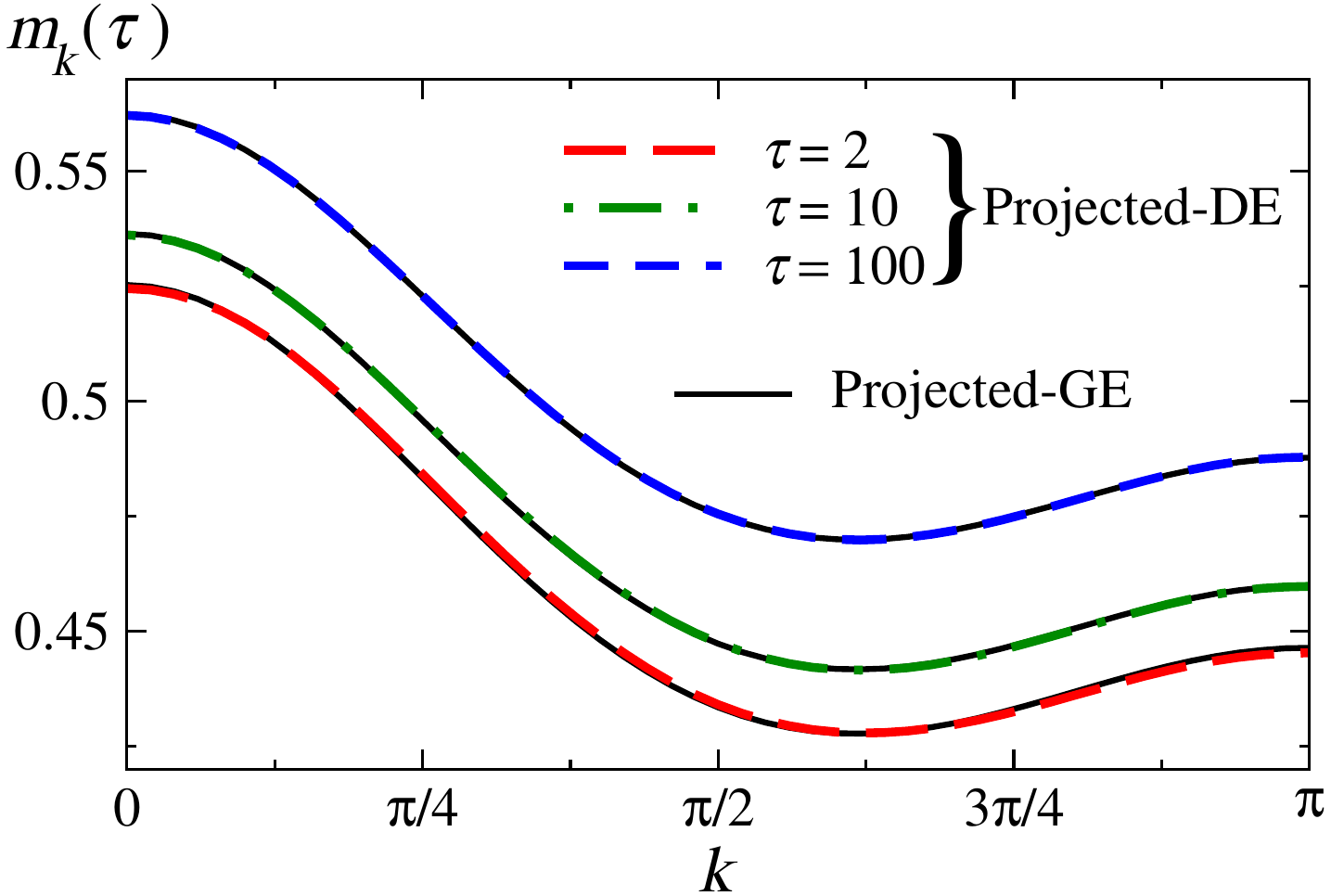}
\vspace{-0.1cm}
\caption{Projected dynamics of $m_k(\tau)$ in the basis of $\hat{H}_0$ at three times after the quench $\hat{H}_I\rightarrow\hat{H}_1$, with $g_1=0.12$. We compare results for the projected dynamics within the diagonal ensemble (Projected DE), also reported in Fig.~\ref{mom_dist}, with those obtained within the grand canonical ensemble (Projected GE). The temperature and the chemical potential of the grand canonical ensemble are set so that the energy density $e_0$ and the particle filling $n$ in this ensemble match those in the diagonal ensemble. The results reported are for the 16th order of the NLCE.}\label{mom_dist_thermal}
\end{figure}

We should stress that all our numerical results for the projected dynamics were obtained within the diagonal ensemble, which allowed us to study indistinctively quantum chaotic reference dynamics in Sec.~\ref{sec:num}, and strongly interacting integrable reference dynamics in Fig.~\ref{part_num_integrable}. However, because of eigenstate thermalization in quantum chaotic systems, and its generalized version in integrable systems, one can equally well use traditional ensembles of statistical mechanics for the projected dynamics when the reference dynamics is quantum chaotic~\cite{dalessio_kafri_16} and generalized Gibbs ensembles when the reference dynamics is integrable~\cite{vidmar_rigol_16}. As an example, in Fig.~\ref{mom_dist_thermal} we compare the results for the projected dynamics in the diagonal ensemble reported in Fig.~\ref{mom_dist} for $g_1=0.12$ at times $\tau=2$, 10, and 100 with those obtained replacing the diagonal ensemble by the grand canonical ensemble with the same energy density $e_0$ and particle filling $n$. As expected, the results obtained within both ensembles are indistinguishable from each other.

\begin{acknowledgements}
M. R. and W. D. R. are grateful to Fabian Essler for insightful discussions over the years about this topic, which motivated this work. We are also grateful to Achim Rosch for valuable comments. We acknowledge support from the National Science Foundation, under Grant No.~PHY-1707482 (K. M. and M. R.), and in part under Grant No.~PHY-1748958. The computations were carried out at the Institute for CyberScience at Penn State.
\end{acknowledgements}

\appendix

\section{Mori-Zwanzig formalism}\label{app:nakajima-zwanzig}

To stress the algebra of the problem, we use a notation different from that in the main text. We consider a linear space $V$ with a projector $\caP: V\to V$, $\caP^2=\caP$, and a linear evolution equation,
\begin{equation}
\dot{a}(\tau)=Ma(\tau),\qquad a(\tau) \in V,\quad M: V\to V.
\end{equation}
Then, we write $\caP_1=\caP$, $\caP_2=(1-\caP)$, $a_i=\caP_ia$, and $M_{ij}=\caP_iM\caP_j$. As the projector is time independent, the evolution equation can be cast as a system of two coupled equations:
\begin{eqnarray}\label{eq:aa1}
\dot{a_1}(\tau) &=&M_{11}a_1(\tau)+M_{12}a_2(\tau), \\\qquad \dot{a_2}(\tau)&=& M_{21}a_1(\tau)+M_{22}a_2(\tau).
\end{eqnarray}
We first formally solve the second of these equations,
\begin{equation}\label{eq:app12}
a_2(\tau)= e^{\tau M_{22}}a_2(0)+ \int_0^\tau \text{d} s\, e^{s M_{22}} M_{21}a_1(\tau-s),
\end{equation}
and insert Eq.~\eqref{eq:app12} into Eq.~\eqref{eq:aa1}, yielding
\begin{eqnarray}
\dot{a_1}(\tau)&=& M_{11}a_1(\tau)+M_{12}e^{\tau M_{22}}a_2(0) \nonumber\\
 && + \int_0^\tau \text{d} s M_{12}  e^{s M_{22}} M_{21}a_1(\tau-s) .
\end{eqnarray}
See Eqs.~\eqref{eq:eomforp} and~\eqref{eq:eomforbarp} in the main text. The relevance of this framework to irreversible phenomena was noticed in Refs.~\cite{zwanzig1960ensemble, mori1965transport}.

\section{From $\caK$ to the autonomous equation} \label{app:fgrfromcak}

Here we look into the superoperator $\caK$ [see Eq.~\eqref{eq:caK}],
\begin{equation}
\caK= \int_0^{+\infty} \text{d}s \caP\caL_1 \e^{s\caL_0}\caL_1\caP,
\end{equation}
in more detail. Each $\caL_1$ contains a commutator $-ig[\hat V,\cdot]$, so in total there are four terms, depending on whether $g\hat V$ in each $\caL_1$ term acts on the left or on the right. We decompose
\begin{equation}
\caK= \caK_{\mathrm{gain}}+\caK_{\mathrm{loss}},
\end{equation}
where the ``gain'' operator contains the two cases in which the $\hat V$'s act on different sides, and the ``loss'' operator contains the two cases in which they act on the same side. 

For both operators, we can recast the two cases  into a single formula by extending the integration range of $s$ from $[0,\infty)$ to $(-\infty,\infty)$. One then has that
\begin{equation}\label{eq:expressiongain}
\caK_{\mathrm{gain}}\hat \rho=g^2\int_0^\infty \text{d} e_0 \text{d} q \hat \rho_{e_0,q} \int_{-\infty}^\infty \text{d} s\, \Tr \left[\hat P_{e_0,q} \hat V_0(s)\caP \hat \rho \hat V\right],
\end{equation}
while for $\caK_{\mathrm{loss}}$ acting on $\hat \rho$ one has
\begin{equation}\label{eq:expressionloss} 
\caK_{\mathrm{loss}}\hat \rho =- g^2\int \text{d} e \text{d} q \hat \rho_{e_0,q}  \int_{-\infty}^\infty \text{d}s \Tr \left[\hat P_{e_0,q} \hat V_0(s) \hat V \caP \hat \rho \right],
\end{equation}
where we abbreviate $\hat P_{e_0,q}= P(\hat H_0 \approx {e}_0L) P( \hat Q \approx {q} L) $ (see Sec.~\ref{sec:slowvariables}), and we use the invariance of $\caP\hat\rho$ and $\hat P_{e_0,q}$ under the evolution generated by $\hat H_0$. Note that we can assume, without loss of generality, that $\hat V$ satisfies $\langle \hat V \rangle_{e_0,q}=0$ for all $(e_0,q)$, because a term with nonzero mean would have canceled out in the commutators [alternatively, it would cancel between Eqs.~\eqref{eq:expressiongain} and~\eqref{eq:expressionloss}].  For this reason, we can replace all correlation functions below by truncated correlation functions. 

Recall that we identified the space of $\caP\hat\rho$ with distributions $p$ on $(e_0,q)$. Indeed, any $\caP\hat\rho$ is of the form $\hat\rho_p$ with distribution $p$; see Eq.~\eqref{eq:densityfromdistribution}. Therefore, we can now abuse notation and interpret $\caK$ as a kernel on the space of distributions $p$. We then have 
\begin{equation}
\caK p(e_0',q')=\int \text{d} e_0 \text{d} q \caK(e_0,q;e_0',q') p(e_0,q),
\end{equation}
and, from the above formulas, we identify
\begin{eqnarray}
\caK(e_0,q;e_0',q') &=& g^2\int_{-\infty}^\infty \text{d} s\, \Tr\left[\hat P_{e'_0,q'} \hat V_0(s)\hat\rho_{e_0,q} \hat  V\right] \label{eq:explicitkernel} \\ &&- g^2\int_{-\infty}^\infty \text{d}s\,  \Tr \left[\hat P_{e'_0,q'} \hat V_0(s) \hat  V \hat \rho_{e_0,q}\right]. \nonumber
\end{eqnarray}

To unravel this further, we introduce
\begin{equation}\label{eq:deltaq}
\hat V^{\delta Q} = \sum_{Q,Q'} \delta_{Q',Q+\delta Q} \hat P_{Q'} \hat V \hat P_{Q},
\end{equation}
where $\hat P_{Q}$ are spectral projections of $\hat Q$ and the sum runs over the eigenvalues of $\hat Q$. Note that the admissible values of $\delta Q$ are $\caO(1)$ because $\hat{V}$ is a sum of $\caO(1)$ local terms. For natural examples of $\hat{Q}$, such as the total particle number operator, we see that also $\hat V^{\delta Q}$ is a sum of local terms. In that case Eq.~\eqref{eq:deltaq} is manifestly of order $\caO(L)$, as a truncated correlation function in equilibrium. The first term in Eq.~\eqref{eq:explicitkernel} can be written as
\begin{equation}\label{eq:definitionr}
r_{e_0,q}(\delta Q) = g^2\int_{-\infty}^\infty \text{d} s\, \Tr\left[ \hat V^{\delta Q}_0(s)\hat\rho_{e_0,q} \hat  V^{-\delta Q}\right],
\end{equation}
with $e'_0=e_0$ and $ q'-q=\delta Q/L$. Indeed, the condition $e_0'=e_0$ is enforced by the integral over $s$. The second term in Eq.~\eqref{eq:explicitkernel} has a contribution at $e_0=e'_0$, $q=q'$ only (by the cyclic property of the trace, the projector $\hat P_{e'_0,q'}$ is put next to the density matrix $\hat \rho_{e_0,q}$) and its value is 
\begin{equation}
\sum_{\delta Q}r_{e_0,q}(\delta Q).
\end{equation}
One could also have guessed this value because the process generated by $\caK$ conserves probability, which translates to
\begin{equation}
\int \text{d}q \text{d}e'_0\caK(e_0,q;e_0',q') =0.
\end{equation}
We have now explicitly interpreted the process generated by $\caK$ as a jump process, with jump rates $r_{e_0,q}(\delta Q) \geq 0$ for jumps in the density $q$ of order $1/L$. The link to the ``drift'' computed in Sec.~\ref{sec:autonomousequation} is by 
\begin{equation}
d(e_0,q) = \sum_{\delta Q} \delta Q \frac{1}{L} r_{e_0,q}(\delta Q) 
\end{equation}
The expressions in Sec.~\ref{sec:autonomousequation} are recovered by writing Eq.~\eqref{eq:definitionr} in terms of eigenkets of $\hat H_0$, and this eventually yields Eq.~\eqref{eq:evolutionondensities}.

\section{Convergence of NLCE and exact diagonalization}\label{NLCEvsED}

All the numerical results reported in the main text, but the relaxation rates $\Gamma^\text{Fermi}$ computed using Fermi's golden rule and full exact diagonalization, were obtained using NLCE calculations. The basics of NLCEs was summarized in Sec.~\ref{sec:nlce}, and relevant parameters for the NLCE calculations (orders, largest Hilbert spaces involved, etc) were mentioned in Sec.~\ref{sec:calcparameters}. Here we discuss the convergence of the NLCE calculations and finite-size effects in the full exact diagonalization calculations.

\begin{figure}[!t]
\includegraphics[width=0.9\columnwidth]{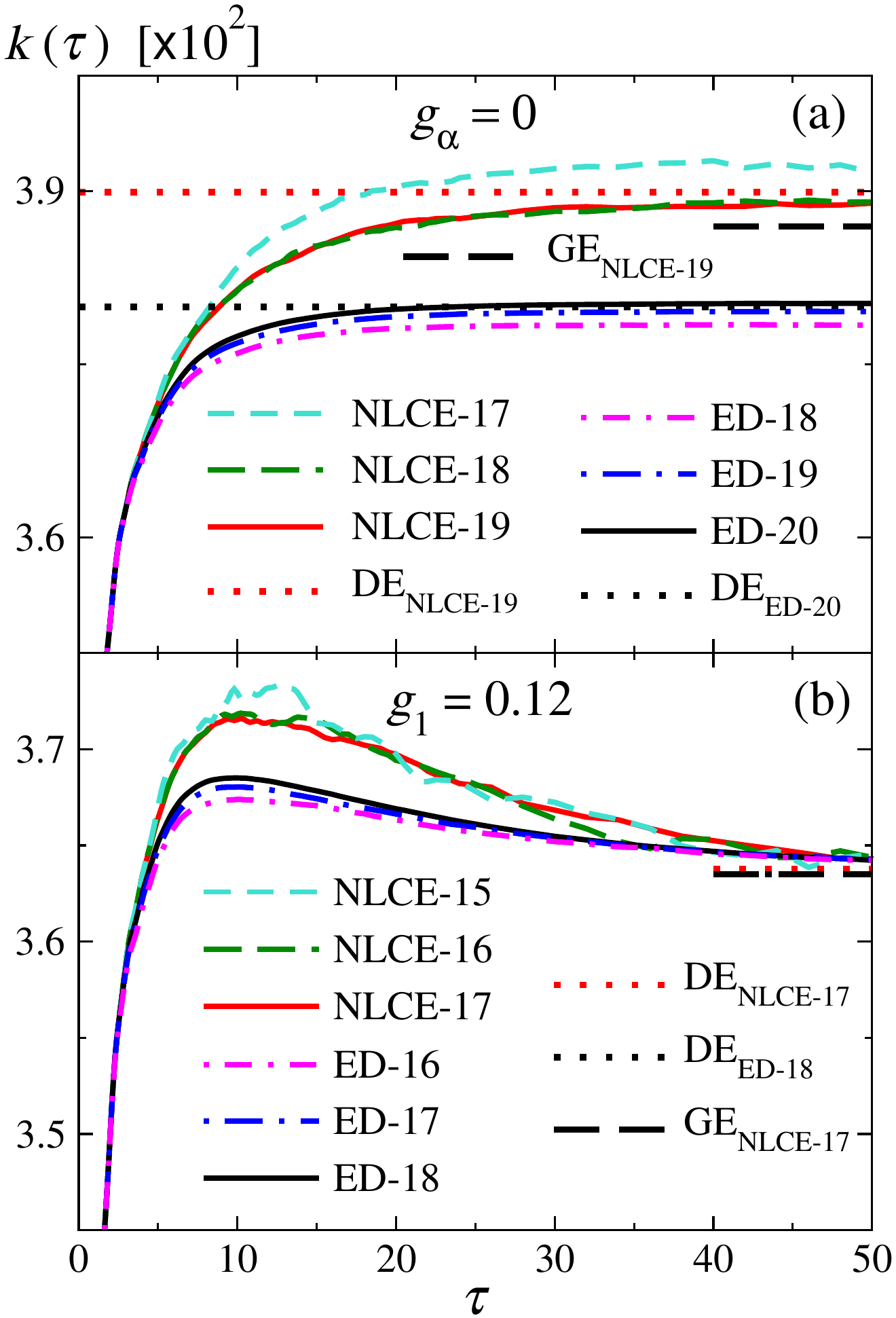}
\vspace{-0.1cm}
\caption{Time evolution of the one-body nearest neighbor correlation $k(\tau)$ obtained in the last three orders $l$ of the NLCE (NLCE-$l$ in the legends), and in the three largest chains with $L$ sites and periodic boundary conditions diagonalized (ED-$L$ in the legends). Results for the diagonal ensemble are shown for the last order of the NLCE (DE$_{\text{NLCE-}l}$) and the largest chain diagonalized (DE$_{\text{ED-}L}$), while results for the grand canonical ensemble are shown for the last order of the NLCE (GE$_{\text{NLCE-}l}$), which are converged to machine precision. (a) Quenches $\hat{H}_I\rightarrow\hat{H}_0$, for $l=17$, 18, and 19 (NLCE-17, NLCE-18, and NLCE-19), and for $L=18$, 19, and 20 (ED-18, ED-19, and ED-20), along with the diagonal ensembles for $l=19$ (DE\ts{NLCE-19}) and $L=20$ (DE\ts{ED-20}), and the grand canonical ensemble for $l=19$ (GE\ts{NLCE-19}). (b) Quenches $\hat{H}_I\rightarrow\hat{H}_1$, in which $g_1=0.12$, for $l=15$, 16, 17 (NLCE-15, NLCE-16, NLCE-17), and for $L=17$, 18, 19 (ED-17, ED-18, ED-19), along with the diagonal ensembles for $l=17$ (DE\ts{NLCE-17}) and $L=18$ (DE\ts{ED-18}), and the grand canonical ensemble for $l=17$ (GE\ts{NLCE-17}). Higher orders in the NLCE, and larger chains in exact diagonalization, are calculated in (a) than in (b) thanks to particle-number conservation in the former.}\label{fig_Nhopp_NLCE_vs_ED}
\end{figure}

All our full exact diagonalization calculations are carried out in chains with periodic boundary conditions. We use translational symmetry to block diagonalize the Hamiltonian, which allows us to study larger chains than within the NLCE calculations. In the exact diagonalization calculations when $g_\alpha\neq0$, in the absence of particle-number conservation, the largest Hamiltonian sector diagonalized has 14\,602 states (for $L=18$). When $g_\alpha=0$, in the presence of particle-number conservation, the largest Hamiltonian sector diagonalized has 9252 states (for $L=20$). We only report exact diagonalization results for $g_\alpha=0$ in Fig.~\ref{fig_Nhopp_NLCE_vs_ED}(a). The matrices involved in our full exact diagonalization calculations are complex for sectors with total quasimomentum $k\neq0$ and $\pi$, and the results reported contain the contribution from all $L$ quasimomentum sectors.

\begin{figure}[!t]
\includegraphics[width=0.985\columnwidth]{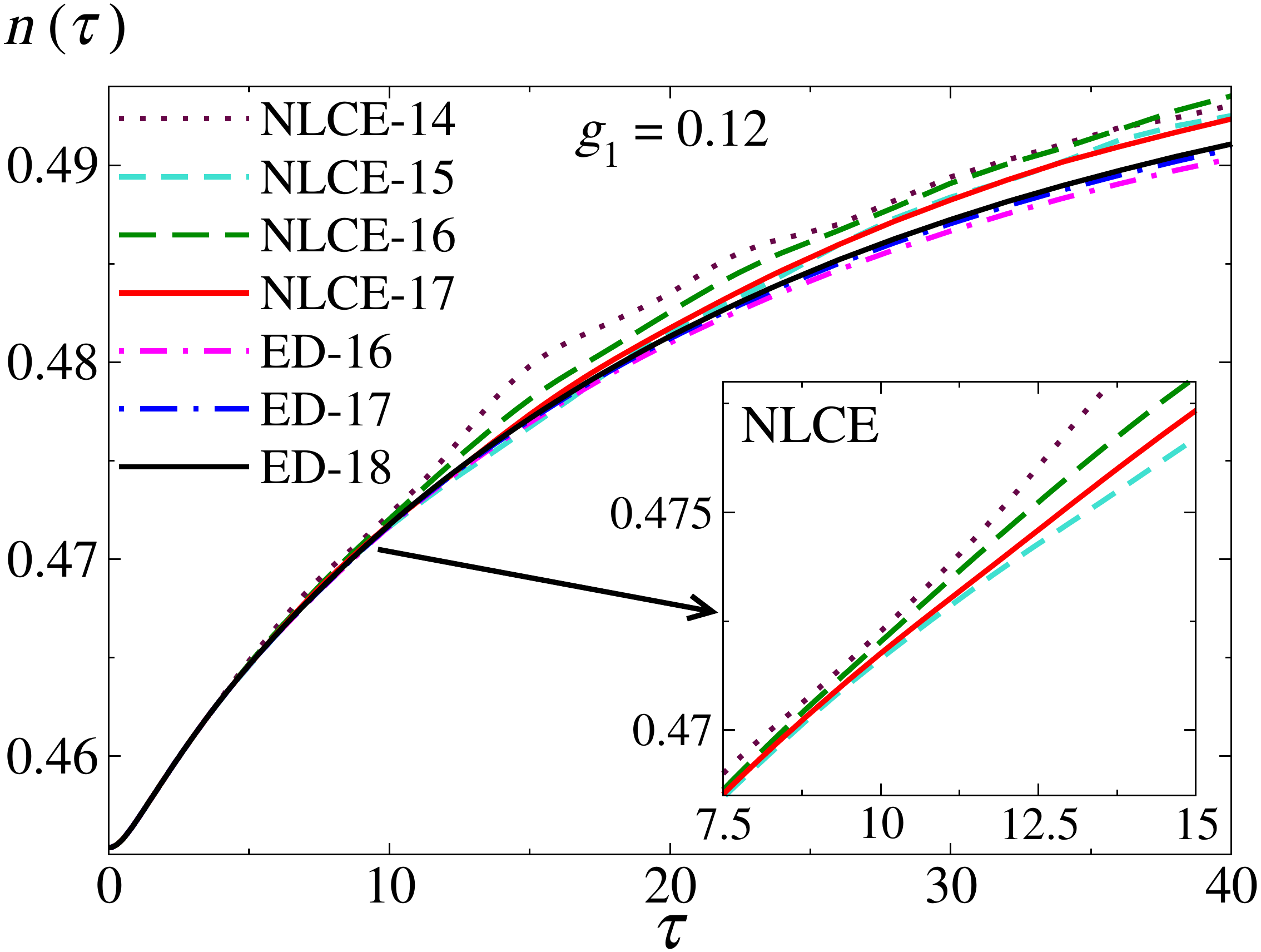}
\vspace{-0.1cm}
\caption{Time evolution of the filling $n(\tau)$, for $g_1=0.12$, obtained in the last four orders $l$ of the NLCE (NLCE-$l$ in the legends), and in the three largest chains with $L$ sites and periodic boundary conditions solved using full exact diagonalization (ED-$L$ in the legends). The inset shows an enlargement of the NLCE results between $\tau=7.5$ and 15.}\label{part_num_NLCE_ED}
\end{figure} 

In Fig.~\ref{part_num_NLCE_ED}, we show the evolution of the particle filling $n(\tau)$ for $g_1=0.12$ (the fastest changing case studied) as obtained in the last four orders of the NLCE, and in the three largest chains solved using full exact diagonalization. In the scale of the figure, all the results are nearly indistinguishable from each other up to $\tau\approx10$. Beyond that time, but not too far from it, the NLCE results can be seen to oscillate in the order of the expansion, even orders of the NLCE are above the odd orders (see the inset in Fig.~\ref{part_num_NLCE_ED}). With increasing the order of the NLCE, one finds that the amplitude of the oscillation decreases and the results converge at longer times. This is similar to what happens in thermal equilibrium calculations, in which the NLCEs converge at lower temperatures as one increases the order of the expansion~\cite{rigol2006numerical, *rigol2007numerical1}. The exact diagonalization results, on the other hand, can be seen to approach the NLCE ones monotonically with increasing the chain size. In Fig.~\ref{part_num_NLCE_ED}, the $n(\tau)$ results in the last order of the NLCE and in the largest periodic chain diagonalized are nearly indistinguishable up to $\tau\approx20$. 

To gain a more quantitative understanding of the convergence of the particle filling $n_l(\tau)$ calculations within NLCE, where $l$ is the order of the expansion, and of finite-size effects in the exact diagonalization (ED) calculations of $n_L(\tau)$, where $L$ is the chain size, we compute the relative differences,
\begin{equation}
\Delta^\text{NLCE}_{l}\left[n(\tau)\right]=
\dfrac{|n_{l=17}(\tau)-n_l(\tau)|}{|n_I-1/2|}\label{eq:delta_n_NLCE}
\end{equation}
between the NLCE results at order $l$ and the last order calculated $l=17$, and  
\begin{equation}
\Delta^\text{ED}_{L}\left[n(\tau)\right]=
\dfrac{|n_{L=18}(\tau)-n_L(\tau)|}{|n_I-1/2|}\label{eq:delta_n_ED}
\end{equation}
between the results for chains with $L$ sites and the largest chain $L=18$ diagonalized. We also compute the relative differences between the last order of the NLCE and the largest periodic chain diagonalized:
\begin{equation}
\Delta^\text{NLCE-ED}\left[n(\tau)\right]=
\dfrac{|n_{l=17}(\tau)-n_{L=18}(\tau)|}{|n_I-1/2|}\label{eq:delta_n_ED_NLCE}.
\end{equation}
In Eqs.~\eqref{eq:delta_n_NLCE}--\eqref{eq:delta_n_ED_NLCE}, $n_I$ is the initial particle filling, which is obtained within machine precision at the 17th order of the NLCE, and 1/2 is the filling in the diagonal ensemble.

\begin{figure}[!t]
\includegraphics[width=0.985\columnwidth]{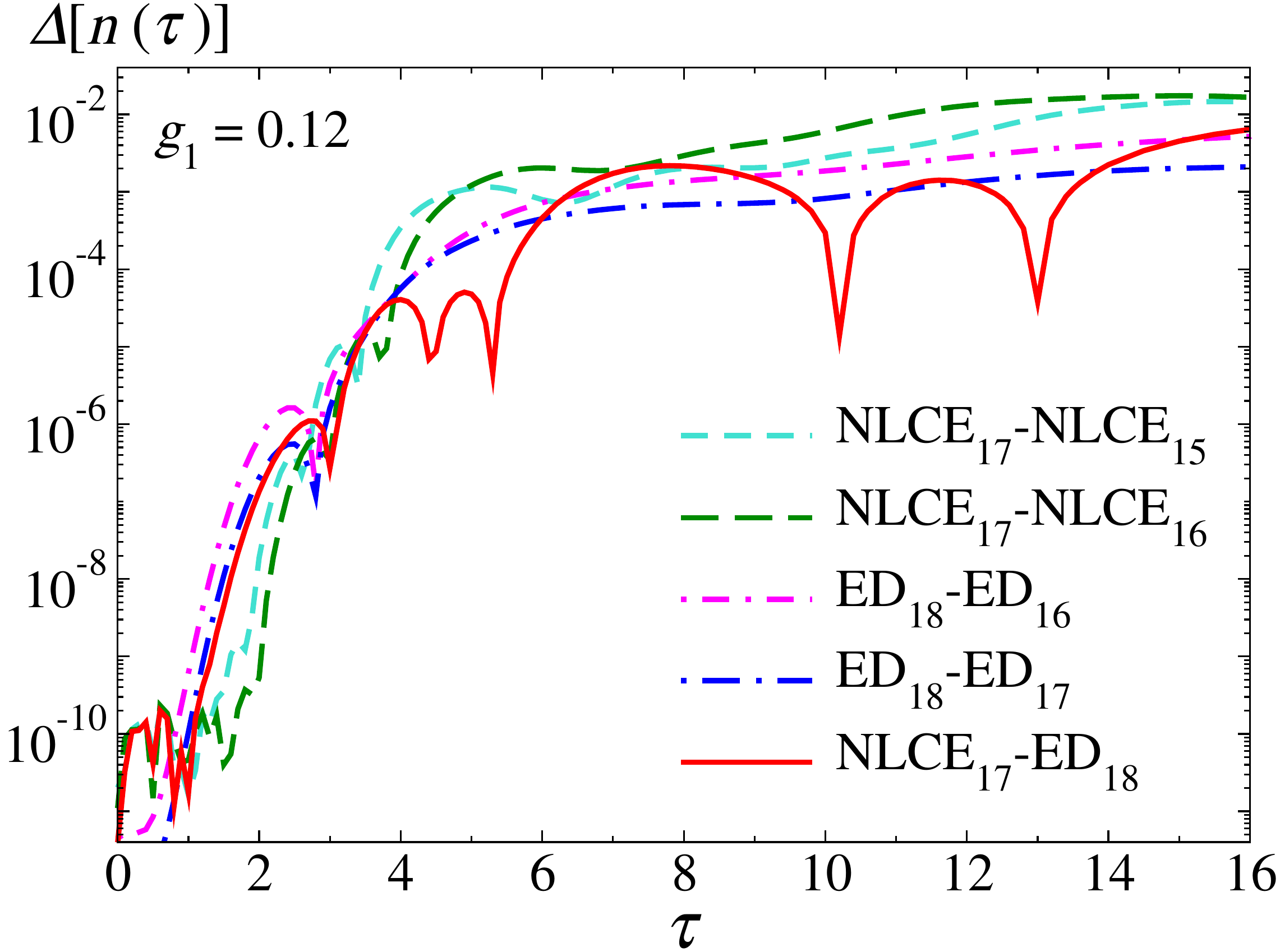}
\vspace{-0.1cm}
\caption{Relative differences, defined in Eqs.~\eqref{eq:delta_n_NLCE}--\eqref{eq:delta_n_ED_NLCE}, for $g_1=0.12$. The exact diagonalization results are obtained in chains with periodic boundary conditions.}\label{part_num_errors_NLCE_ED}
\end{figure}

In Fig.~\ref{part_num_errors_NLCE_ED}, we plot the relative differences defined in Eqs.~\eqref{eq:delta_n_NLCE} and~\eqref{eq:delta_n_ED} for $l=15$ and 16 of the NLCE, and for $L=16$ and 17 in the exact diagonalization calculations, also for $g_1=0.12$ as in Fig.~\ref{part_num_NLCE_ED}. Some points to be highlighted from the plots in Fig.~\ref{part_num_errors_NLCE_ED} are the following: (i) the NLCE results for $l=16$ are likely converged within machine precision up to $\tau\approx2$, while the exact diagonalization ones for $L=17$ are likely converged within machine precision only up to about one half of that time ($\tau\approx1$); (ii) the relative differences between the various orders of the NLCE and between the various exact diagonalization calculations are below 0.01\% for $\tau\lesssim4$; (iii) at times $\tau\approx16$, the relative differences are in all cases below 2\% (they are smaller between the exact diagonalization results than between the NLCE ones). 

The results for the relative difference $\Delta^\text{NLCE-ED}\left[n(\tau)\right]$ are qualitatively similar to those for $\Delta^\text{NLCE}_{l}\left[n(\tau)\right]$ and $\Delta^\text{ED}_{L}\left[n(\tau)\right]$. We find that $\Delta^\text{NLCE-ED}\left[n(\tau)\right]\leq0.01\%$ for times $\tau\leq5$. We use times $\tau\leq5$ for the exact diagonalization calculation of the rates from Fermi's golden rule in Appendix~\ref{fermi_app}. We also find that $\Delta^\text{NLCE-ED}\left[n(\tau)\right]\lesssim0.5\%$ for times $\tau\leq16$. We use times $\tau\leq16$ in the fits to obtain the rates from the 17th order of the NLCE dynamics in Sec.~\ref{sec:nearhopp}.

Next, for the one-body nearest neighbor correlations $k(\tau)$, we discuss the convergence of the NLCE calculations $k_l(\tau)$ and finite-size effects in the full exact diagonalization calculations $k_L(\tau)$. $k(\tau)$, being a correlation function, is more challenging to obtain accurately than $n(\tau)$, which is a thermodynamic quantity. 

In Fig.~\ref{fig_Nhopp_NLCE_vs_ED}(a), we show dynamics after the quench $\hat{H}_I\rightarrow \hat{H}_0$ for $k_l(\tau)$ with $l=$ 17, 18, and 19 (the latter two are also reported in Fig.~\ref{hopping_unperturbed}) and for $k_L(\tau)$ with $L$ = 18, 19, and 20. $k_l(\tau)$ and $k_L(\tau)$ are almost indistinguishable from each other up to times $\tau\approx5$ (the earliest times are not shown to gain dynamical range in the $y$ axis). For $\tau\gtrsim5$, the NLCE and exact diagonalization results split and each equilibrate to the prediction of the corresponding diagonal ensemble, $k_{l=19}^\text{DE}$ (DE\ts{NLCE-19}) and $k_{L=20}^\text{DE}$ (DE\ts{ED-20}), respectively. As in Ref.~\cite{mallayya2018quantum}, the NLCE results for the diagonal ensemble (DE\ts{NLCE-19}) are closer to the grand canonical ensemble ones $k_{l=19}^\text{GE}$ (GE\ts{NLCE-19}) than the exact diagonalization results for the diagonal ensemble (DE\ts{ED-20}). The differences between them are due to lack of convergence of the diagonal ensemble within NLCEs and finite-size effects for the diagonal ensemble within exact diagonalization. The NLCE results for the grand canonical ensemble are converged within machine precision. Consequently, at long times, the NLCE is expected to be more accurate than exact diagonalization. At intermediate times $5\lesssim\tau\lesssim15$, $k_{l=16}(\tau)$ and $k_{l=17}(\tau)$ are almost indistinguishable from each other, while $k_{L}(\tau)$ shifts upward with increasing $L$, toward the NLCE predictions. This suggests that NLCE is also more accurate than exact diagonalization at intermediate times.

In Fig.~\ref{fig_Nhopp_NLCE_vs_ED}(b), we show dynamics after the quench $\hat{H}_I\rightarrow \hat{H}_1$ ($g_1=0.12$) for $k_l(\tau)$ with $l=$ 15, 16, and 17 (the latter two are also reported in Fig.~\ref{hopping_perturbed}) and for $k_L(\tau)$ with $L$ = 16, 17, and 18. $k_l(\tau)$ and $k_L(\tau)$ are almost indistinguishable from each other up to times $\tau\approx4$ (again, the earliest times are not shown to gain dynamical range in the $y$ axis). For $\tau\gtrsim4$, the NLCE and exact diagonalization results again split, as for the quench $\hat{H}_I\rightarrow \hat{H}_0$, despite the fact that both equilibrate to diagonal ensemble results that are very close to each other. $k_{l=17}^\text{DE}$ (DE\ts{NLCE-17}) and $k_{L=18}^\text{DE}$ (DE\ts{ED-18}) in Fig.~\ref{fig_Nhopp_NLCE_vs_ED}(b) are almost indistinguishable from each other and from the grand canonical ensemble result $k_{l=17}^\text{GE}$ (GE\ts{NLCE-17}). At times $5\lesssim\tau\lesssim15$, $k_{l=16}(\tau)$ and $k_{l=17}(\tau)$ are almost indistinguishable, while $k_{L}(\tau)$ shifts upward toward the NLCE predictions with increasing $L$. As for the quenches $\hat{H}_I\rightarrow \hat{H}_0$, these results suggest that NLCE is more accurate than exact diagonalization at intermediate times in quenches $\hat{H}_I\rightarrow \hat{H}_1$. 

The discrepancy between the NLCE and exact diagonalization results at intermediate times after quenches $\hat{H}_I\rightarrow \hat{H}_1$ is likely a manifestation of finite-size effects in $k_L(\tau)$ that result from the ``prethermal'' dynamics seen in Fig.~\ref{fig_Nhopp_NLCE_vs_ED}(a), which equilibrates to a diagonal ensemble value ($k_{L=20}^\text{DE}$) that is lower than the expected grand canonical ensemble one. The results in Fig.~\ref{fig_Nhopp_NLCE_vs_ED}(b) also show that, because of finite-size effects, the thermalization rates for $k(\tau)$ are smaller in exact diagonalization, and increasing with increasing $L$, than in NLCE.

\section{Thermalization rates from Fermi's golden rule}\label{fermi_app}

\begin{figure}[!b]
\includegraphics[width=0.985\columnwidth]{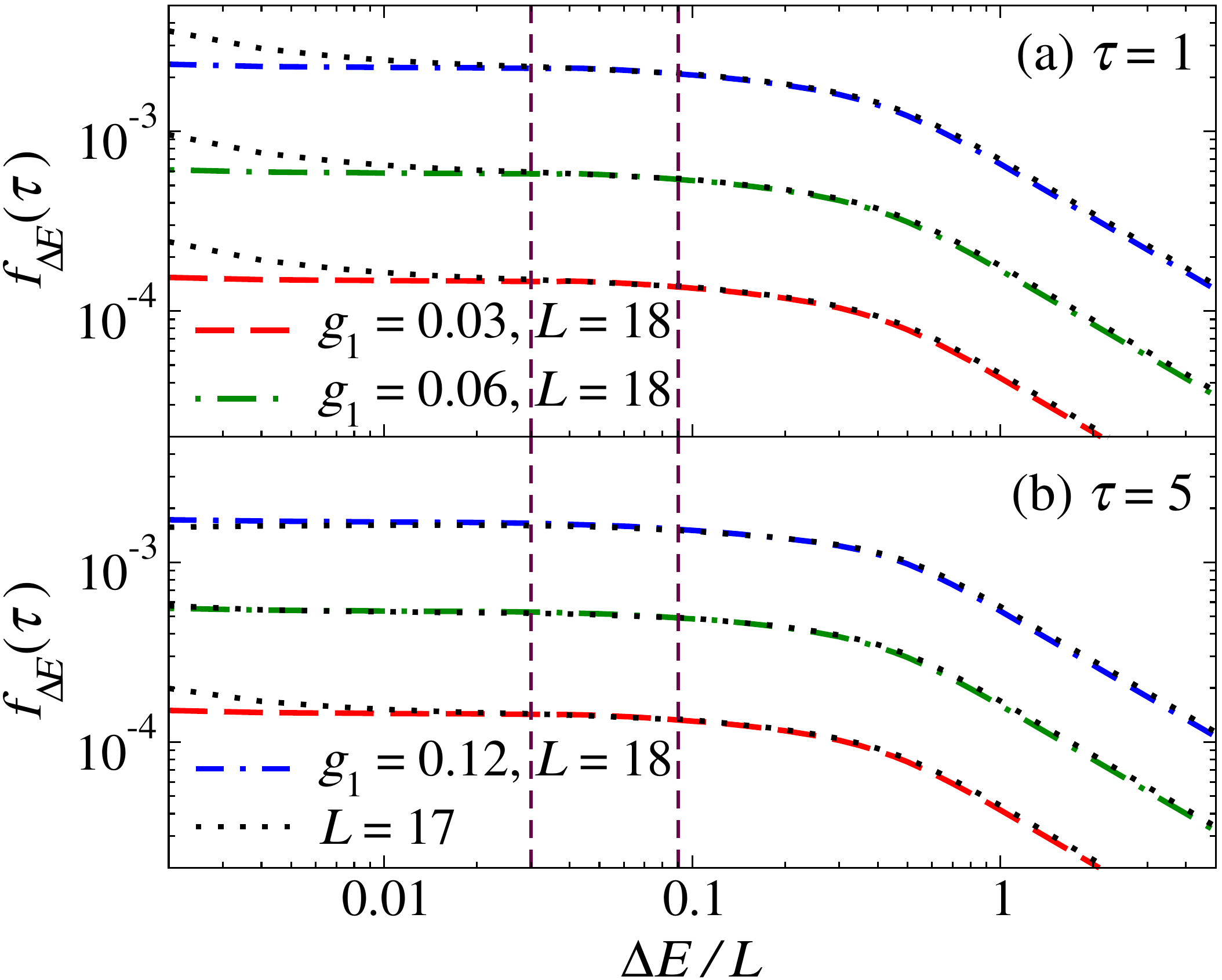}
\vspace{-0.1cm}
\caption{$f_{\Delta E}(\tau)$, see Eq.~\eqref{fermi_rate_dE}, evaluated for a wide range of energy windows $\Delta E$ for quenches $\hat{H}_I\rightarrow\hat{H}_1$ [see Eq.~\eqref{V1}] with $g_1=0.03$, 0.06, and 0.12. $f_{\Delta E}(\tau)$ is computed using full exact diagonalization of chains with $L=18$ (see legends) and $L=17$ (dotted lines) and periodic boundary conditions. The vertical lines delimit the range $\Delta E/L\in[0.03,0.09]$ in which $f_{\Delta E}(\tau)$ is approximately constant.}\label{fermi_dE_plots}
\end{figure}

We use full exact diagonalization in chains with $L$ sites and periodic boundary conditions to evaluate Eq.~\eqref{fermi_rate}. For the numerical calculation, the delta function is replaced by a coarse-graining procedure leading to the following modified version of Eq.~\eqref{fermi_rate}:
\begin{eqnarray}
f_{\Delta E}(\tau) &=&\label{fermi_rate_dE} \dfrac{2\pi g_\alpha^2}{L\Delta E}\sum_{i}P_i(\tau)\\
&&\times\sum_{|E_j-E_i|\le\Delta E/2} |\bra{E^0_j}\hat{V}_\alpha\ket{E^0_i}|^2\left(N_j-N_i\right),\nonumber
\end{eqnarray}
where $\ket{E^0_{i}}$ $(\ket{E^0_j})$ are the eigenkets of $\hat{H}_0$ with energy $E^0_{i}$ ($E^0_j$), $N_i=\bra{E^0_i}\hat{N}\ket{E^0_i}$, and $P_i(\tau)=\bra{E^0_i}\hat{\rho}(\tau)\ket{E^0_i}$.

In Fig.~\ref{fermi_dE_plots}, we plot $f_{\Delta E}(\tau)$, evaluated in chains with $L=17$ and $L=18$ for quenches $\hat{H}_I\rightarrow\hat{H}_1$ [see Eq.~\eqref{V1}], as a function of $\Delta E$. We show results at two times, $\tau=1$ in Fig.~\ref{fermi_dE_plots}(a) and $\tau=5$ in Fig.~\ref{fermi_dE_plots}(b), and for three values of $g_1$ at each time. Our main finding in Fig.~\ref{fermi_dE_plots} is that, in the interval $\Delta E/L\in[0.03,0.09]$, $f_{\Delta E}(\tau)$ is nearly independent of $\Delta E$, and is approximately the same in chains with $L=17$ and $L=18$. Similar results were obtained at times $\tau\in\left[0,5\right]$, for which we showed in Appendix~\ref{NLCEvsED} that our best NLCE and exact diagonalization results for $n(\tau)$ differ by less than $0.01\%$.

Having identified an appropriate range of values of $\Delta E$, we compute the rate
\begin{equation}
\Gamma_{\Delta E,\tau}(g_\alpha)=-\dfrac{f_{\Delta E}(\tau)}{n(\tau)-n_{\text{DE}}},\label{fermi_gamma_dE}
\end{equation}
which is defined following Eq.~\eqref{fermi_gamma}.

\begin{figure}[!t]
\includegraphics[width=0.985\columnwidth]{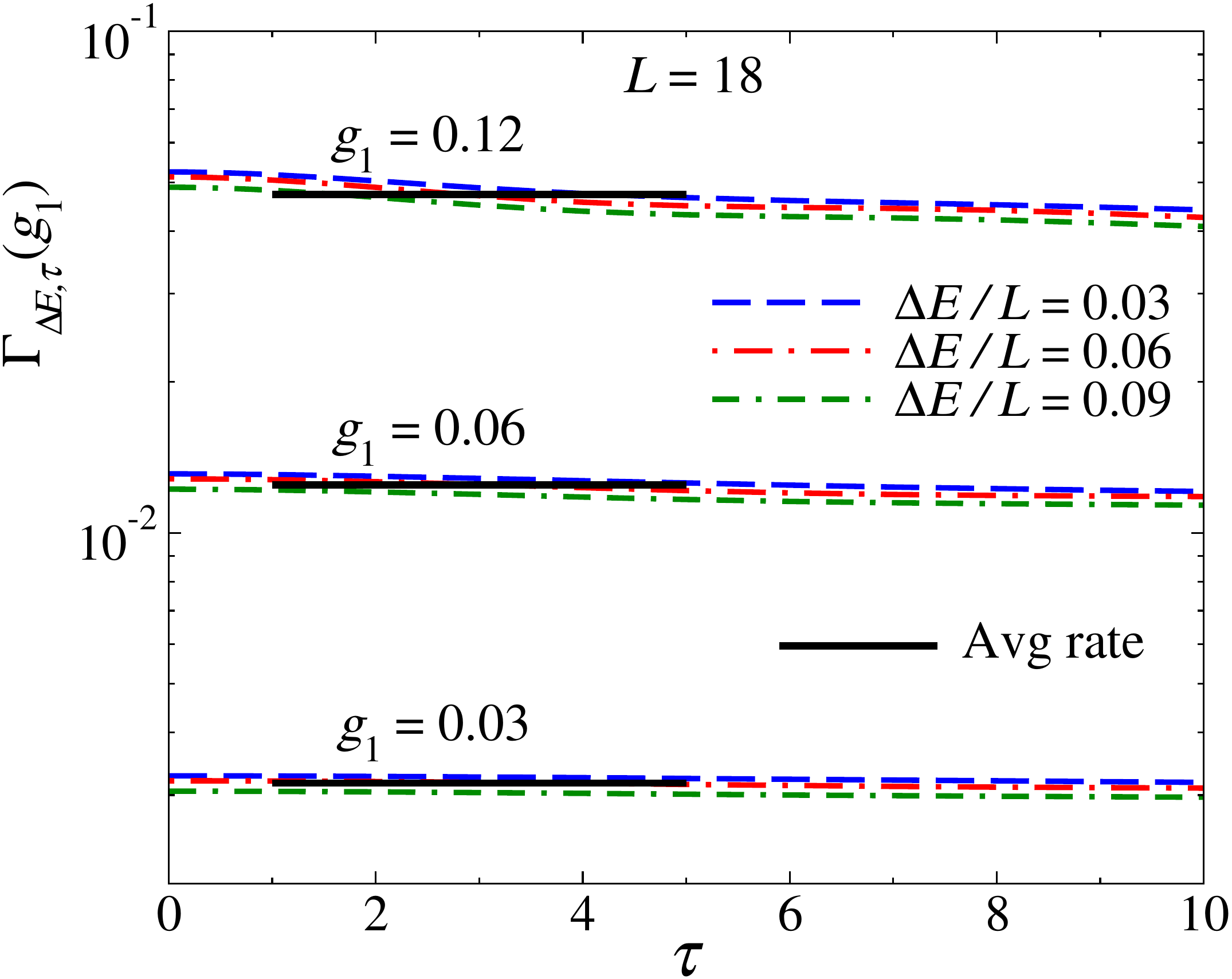}
\vspace{-0.1cm}
\caption{$\Gamma_{\Delta E,\tau}(g_1)$ [see Eq.~\eqref{fermi_gamma_dE}] evaluated for quenches $\hat{H}_I\rightarrow\hat{H}_1$ [see Eq.~\eqref{V1}] with $g_1=0.03$, 0.06, and 0.12. The results reported are from full exact diagonalization of chains with $L=18$ and periodic boundary conditions. The average thermalization rate (horizontal line) is obtained by averaging $\Gamma_{\Delta E,\tau}(g_1)$ over $\tau=1, 1.5, \ldots, 5$ (nine values) and over $\Delta E/L =0.03, 0.032, \ldots, 0.09$ (31 values), for a total of 279 values of $\Gamma_{\Delta E,\tau}(g_1)$ entering each average reported.}\label{fermi_gamma_plot}
\end{figure}

Figure~\ref{fermi_gamma_plot} shows $\Gamma_{\Delta E,\tau}(g_1)$ versus $\tau$ for three values of $g_1$, and for three values of $\Delta E/L$ for each value of $g_1$. The rates for each value of $g_1$ decrease slowly with increasing $\tau$, and are very close to each other for the three values of $\Delta E/L$ shown. Given the results for $n(\tau)$ at times $\tau\leq1$ in Fig.~\ref{fig_Nhopp_NLCE_vs_ED}, which exhibit a plateaulike behavior discussed in Sec.~\ref{sec:partfill}, the rates $\Gamma^\text{Fermi}(g_\alpha)$ for $\tau\leq1$ are not meaningful. Hence, the rates $\Gamma^\text{Fermi}(g_\alpha)$ reported in Sec.~\ref{sec:partfill} were obtained averaging $\Gamma_{\Delta E,\tau}(g_\alpha)$ over the results for $\tau=1, 1.5, \ldots, 5$ (nine values), and for $\Delta E/L =0.03, 0.032, \ldots, 0.09$ (31 values), for a total of 279 values entering each average. In Fig.~\ref{fermi_gamma_plot}, we report the averages as horizontal lines. For each average, we also compute the standard deviation. In Fig.~\ref{part_num_fig}(c), we report the averages, and the standard deviations (as error bars), for different values of $g_1$ and for $L=17$ and $L=18$.

The rates $\Gamma^\text{Fermi}(g_2)$ reported in Fig.~\ref{part_num_integrable} were obtained averaging $\Gamma_{\Delta E,\tau}(g_2)$ over the results for $\tau=1, 1.5, \ldots, 5$ (nine values), and for $\Delta E/L =0.100, 0.102, \ldots, 0.15, 0.16, \ldots, 0.20$ (31 values), for a total of 279 values entering each average.

\bibliography{Reference}
\end{document}